\documentclass[a4paper,11pt]{article}
\usepackage{jinstpub} 
\usepackage{lineno}
\usepackage{subfigure}
\usepackage{graphicx}
\usepackage{dcolumn}
\usepackage{bm}
\usepackage[utf8]{inputenc}
\usepackage{hyperref}
\usepackage{float}
\usepackage{algorithm}
\usepackage{algpseudocode}
\usepackage{diagbox}
\usepackage{booktabs}
\usepackage{makecell}
\usepackage{caption}
\usepackage{multirow}
\usepackage{booktabs}
\usepackage[symbol]{footmisc}

\title{\boldmath Study of residual artificial neural network for particle identification in the CEPC high-granularity calorimeter prototype}

\author[a, b, *]{S.~Song, \note{Corresponding author.}}
\author[a, b]{J.~Chen,}
\author[d, e]{J.~Liu,}
\author[f, g, h]{Y.~Liu,}
\author[f, g, h]{B.~Qi,}
\author[d, e]{Y.~Shi,}
\author[d, e]{J.~Wang,}
\author[c, a, b]{Z.~Wang,}
\author[a, b, c, *]{and H.~Yang}

\affiliation[a]{Institute of Nuclear and Particle Physics, School of Physics and Astronomy, Shanghai Jiao Tong University,\\Dongchuan Road 800, Shanghai 200240, P.R. China}

\affiliation[b]{Key Laboratory for Particle Astrophysics and Cosmology (Ministry of Education), Shanghai Key Laboratory for Particle Physics and Cosmology,\\Dongchuan Road 800, Shanghai 200240, P.R. China}

\affiliation[c]{Tsung-Dao Lee Institute, Shanghai Jiao Tong University,\\Shengrong Road 520, Shanghai 200240, P.R. China}

\affiliation[d]{State Key Laboratory of Particle Detection and Electronics, University of Science and Technology of China,\\Jinzhai Road 96, Hefei 230026, P.R. China}

\affiliation[e]{Department of Modern Physics, University of Science and Technology of China,\\Jinzhai Road 96, Hefei 230026, P.R. China}

\affiliation[f]{Institute of High Energy Physics, Chinese Academy of Sciences (CAS),\\Yuquan Road 19B, Beĳing 100049, P.R. China}

\affiliation[g]{State Key Laboratory of Particle Detection and Electronics, Institute of High Energy Physics, CAS,\\Yuquan Road 19B, Beĳing 100049, P.R. China}

\affiliation[h]{University of Chinese Academy of Sciences,\\East Yanqihu Road 1, Beĳing 100049, P.R. China}

\emailAdd{siyuansong@sjtu.edu.cn, haijun.yang@sjtu.edu.cn}

\abstract{Particle Identification (PID) plays a central role in associating the energy depositions in calorimeter cells with the type of primary particle in a particle flow oriented detector system. In this paper, we propose novel PID methods based on the Residual Network (ResNet) architecture which enable the training of very deep networks, bypass the need to reconstruct feature variables, and ensure the generalization ability among various geometries of detectors, to classify electromagnetic showers and hadronic showers. Using Geant4 simulation samples with energy ranging from 5 GeV to 120 GeV, the efficacy of Residual Connections is validated and the performance of our model is compared with Boosted Decision Trees (BDT) and other pioneering Artificial Neural Network (ANN) approaches. In shower classification, we observe an improvement in background rejection over a wide range of high signal efficiency ($> 95\%$). These findings highlight the prospects of ANN with Residual Blocks for imaging detectors in the PID task of particle physics experiments.
}


\begin{document}
\maketitle
\flushbottom

\section{Introduction}
\label{sec:intro}

The Circular Electron Positron Collider (CEPC), dedicated to precisely measuring the properties of the Higgs boson, incorporates the Particle Flow Algorithm (PFA) \cite{cepc2018cepc} in the detector system's baseline design. The fundamental concept behind the PFA is to utilize the most suitable detector subsystem for accurately determining the energy/momentum of individual particles within a jet. To achieve this objective, PFA-oriented calorimeters require high granularity. In the context of a growing volume of inputs, the development of Particle Identification (PID) methods customized for high-granularity calorimeters holds significance.

\begin{figure}[htbp]
\centering

\subfigure[]
{
 	
    \includegraphics[width=0.35\textwidth]{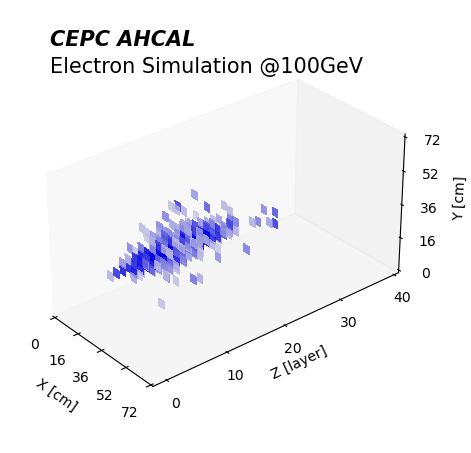}
    
}
\hspace{1cm}
\subfigure[]
{     
    \includegraphics[width=0.35\textwidth]{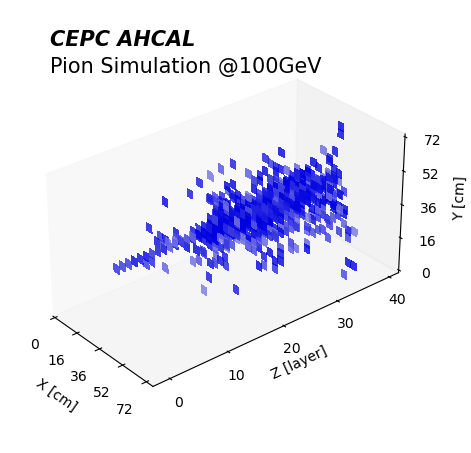}
   
}
\subfigure[]
{
    \centering
    \includegraphics[width=0.35\textwidth]{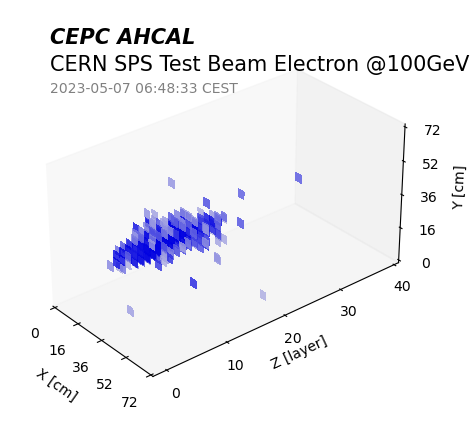}
   
}
\hspace{1cm}
\subfigure[]
{
 	
    \includegraphics[width=0.35\textwidth]{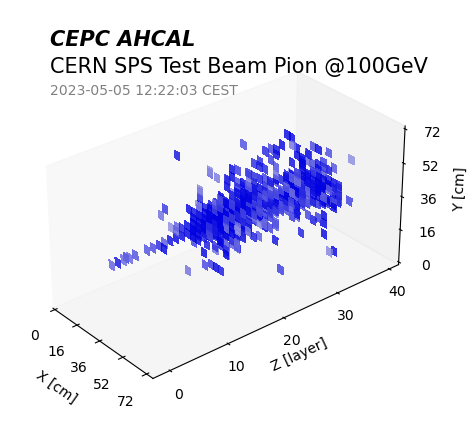}
    
}

\caption{The CEPC Analogue Hadronic Calorimeter (AHCAL) prototype is employed to display different shower types. In the first row, the electron event (a), and the pion event (b) are simulated. In the second row, the electron event (c), and the pion event (d) are selected from the CEPC AHCAL test beam Run files. Their primary energy is 100 GeV.}
\label{fig:shower}
\end{figure}

High-granularity calorimeters capture the intricate spatial development of showers in unprecedented detail, providing valuable information for PID. Patterns of energy depositions observed in the 3D array of ``cells'' substantially reflect the characteristics of the type of the primary particle. Typical spatial configurations of shower types are illustrated in figure \ref{fig:shower}. The electron leads to an electromagnetic shower, and the pion causes a hadronic shower. Distinguishing between electromagnetic and hadronic showers is crucial for gaining insights into the underlying physics processes, while some hadronic showers may contain electromagnetic components, making it difficult to establish clear-cut criteria for classification.

In one way, the Multivariate Analysis (TMVA) technique \cite{hoecker2007tmva} reconstructs shower topology feature variables from such 3D images. For instance, in CALICE SDHCAL PID \cite{liu2020particle}, six shower topology variables were reconstructed to build a TMVA Boosted Decision Trees (BDT) classifier \cite{yang2005studies, roe2005boosted, roe2006boosted} for separating electron events from hadron events. The disadvantage of a BDT is that it heavily depends on the quality and relevance of the reconstructed feature variables. When compressing massive raw data into several feature variables, systematically finding a set of effective variables that are not highly correlated is challenging, and some hidden features might be missed~ \cite{macaluso2018pulling}.

\begin{figure}[htbp]
\centering
    
\includegraphics[width=0.8\textwidth]{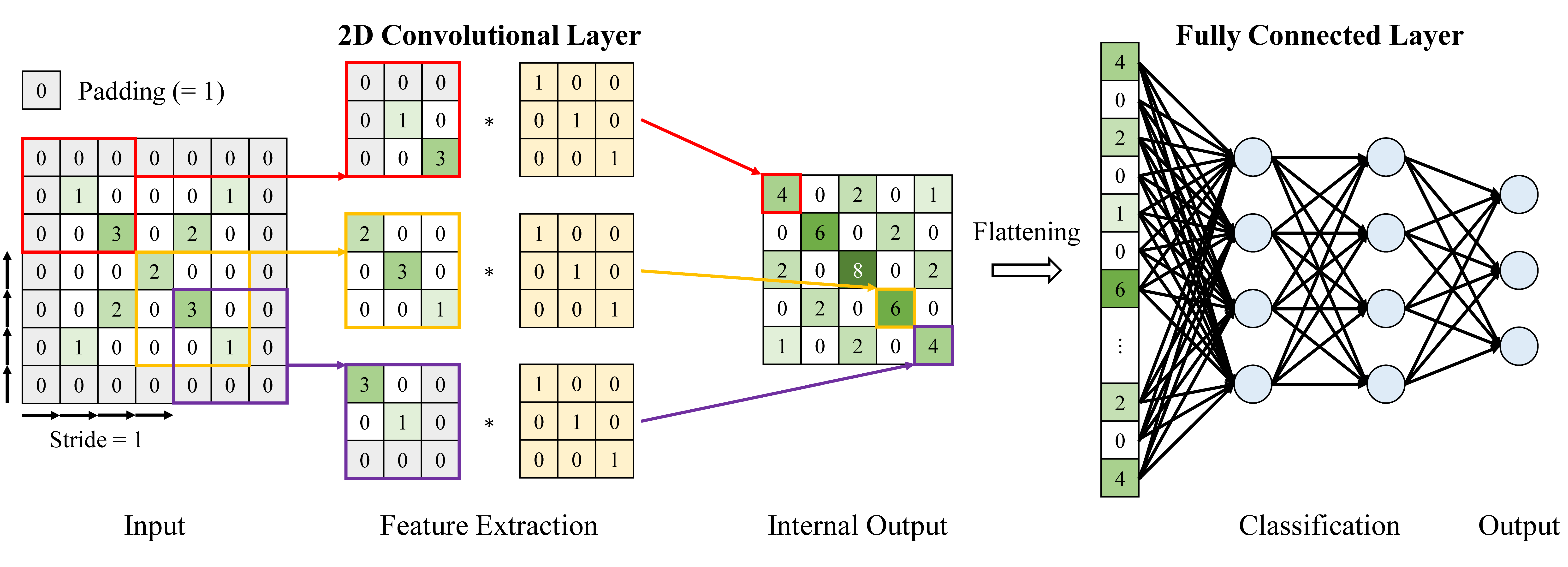}

\caption{The 2D convolutional layers are coupled with fully connected layers. The former is for feature extraction while the latter is for classification.}
\label{fig:cov_fc}
\end{figure}

In another way, Artificial Neural Networks (ANN) have great potential to be applied to high-granularity calorimeter PID \cite{carminati2017calorimetry, bourilkov2019machine}, offering the advantage of making full use of raw high-dimensional input data, (e.g.\ spatial distributions of energy depositions). There are two specialized types of ANN, Convolutional Neural Networks (CNN) designed for processing structured grid-like data, such as images, and Graph Neural Networks (GNN) tailored for processing graph-structured data. Recently, ANN has been utilized in various experiments, including the NOvA experiment for neutrino identification \cite{psihas2020review}, the Belle II experiment for the detection of low-momentum muons and pions \cite{charan2023particle}, and the CMS experiment for the identification of hadronic tau lepton decays \cite{tumasyan2022identification}, etc. In common, these methods share a shallow architecture combining 2D convolutional (Conv) layers for feature extraction and fully connected (FC) layers for classification as illustrated in figure \ref{fig:cov_fc}. Although they have already achieved great success, a specific study of the state-of-the-art deeper architectures such as ResNet\cite{he2016deep} and AlexNet\cite{krizhevsky2012imagenet} on PID has not been conducted. Furthermore, the potential applications of these methods may be limited to detectors with regular geometry, facilitating the preprocessing of data into images. Therefore, there is growing interest in Graph Neural Network (GNN) models, which offer greater flexibility in handling data from detectors with irregular geometries. In this domain, pioneering models such as DGCNN\cite{wang2019dynamic} and GravNet\cite{qasim2019learning, qasim2022end} have been developed for either classification or clustering.

In this paper, we propose novel PID methods based on the ResNet architecture\cite{he2016deep}, a much deeper architecture with millions of trainable parameters tailored for high-granularity calorimeters. The efficacy of introducing Residual Connections is studied and the final performance is compared with the benchmark BDT\cite{chen2016xgboost}, LeNet\cite{lecun1998gradient}, and AlexNet\cite{krizhevsky2012imagenet} methods. Furthermore, to ensure the generality, an updated model adaptable to any irregular detector geometry, called the Dynamic Graph Residual Networks (DGRes), allowing our ResNet-based model to be applied beyond detectors with regular structures, has been proposed and compared with other GNN models, DGCNN\cite{wang2019dynamic} and GravNet\cite{qasim2019learning, qasim2022end}. The paper is structured as follows: Section \ref{sec:dataset} introduces detector geometry and Monte Carlo simulation samples which are used to study the performance of various classifiers, Section \ref{sec:bdt} briefly illustrates PID based on BDT and provides insights on shower topology; Section~\ref{sec:ann} demonstrates the algorithm, the performance, and the complexity of our models; Section \ref{sec:conclusion} concludes this research. For your convenience, the code and samples for this paper are provided.\footnote[2]{code and samples: \href{https://github.com/Tom0126/CEPC-AHCAL-PID}{https://github.com/Tom0126/CEPC-AHCAL-PID}}

\section{Detector geometry and Monte Carlo samples}
\label{sec:dataset}

\begin{figure}[htbp]
\centering
\subfigure[]
{
    \centering
    \includegraphics[width=0.4\textwidth]{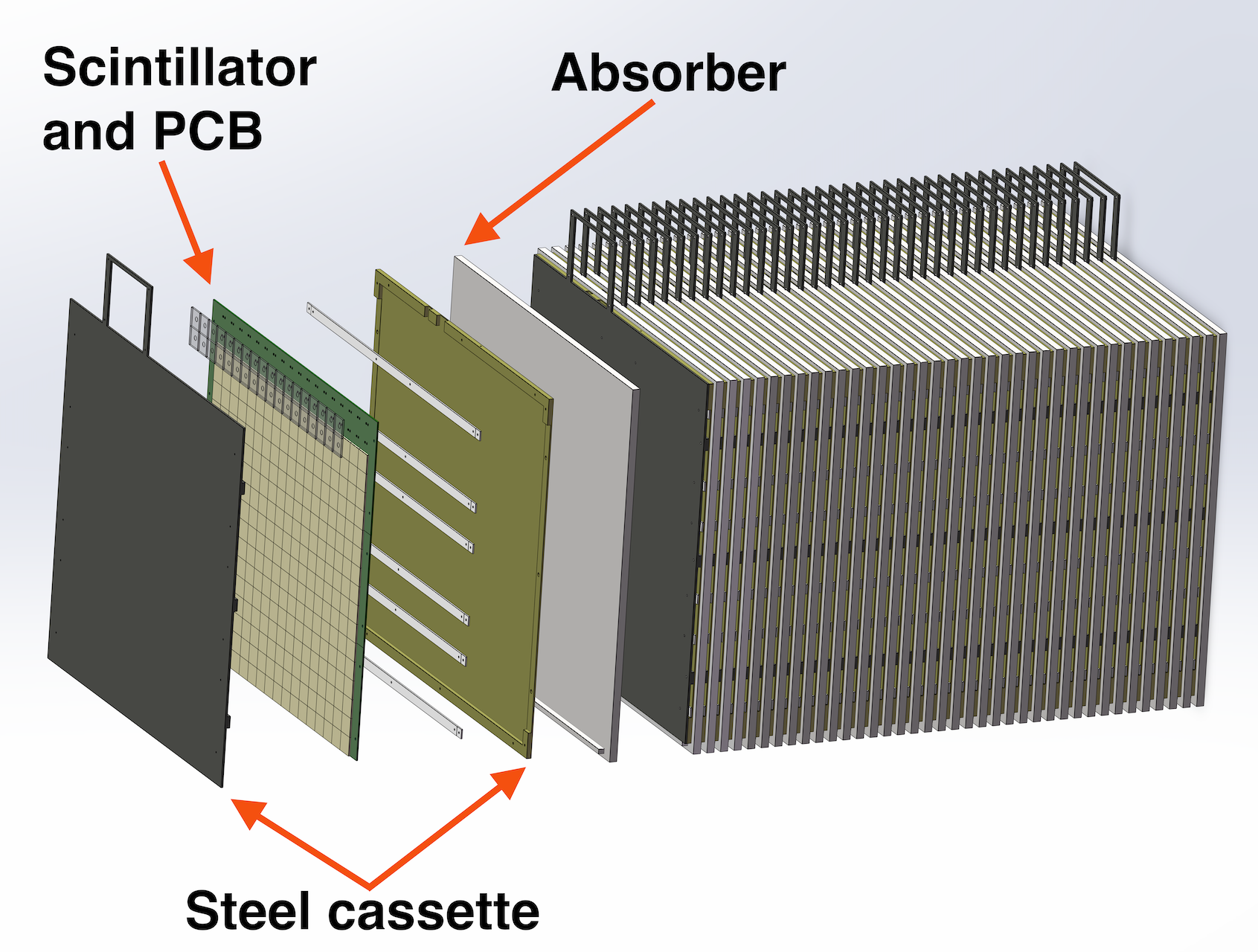}
   
}
\subfigure[]
{     
    \includegraphics[width=0.42\textwidth]{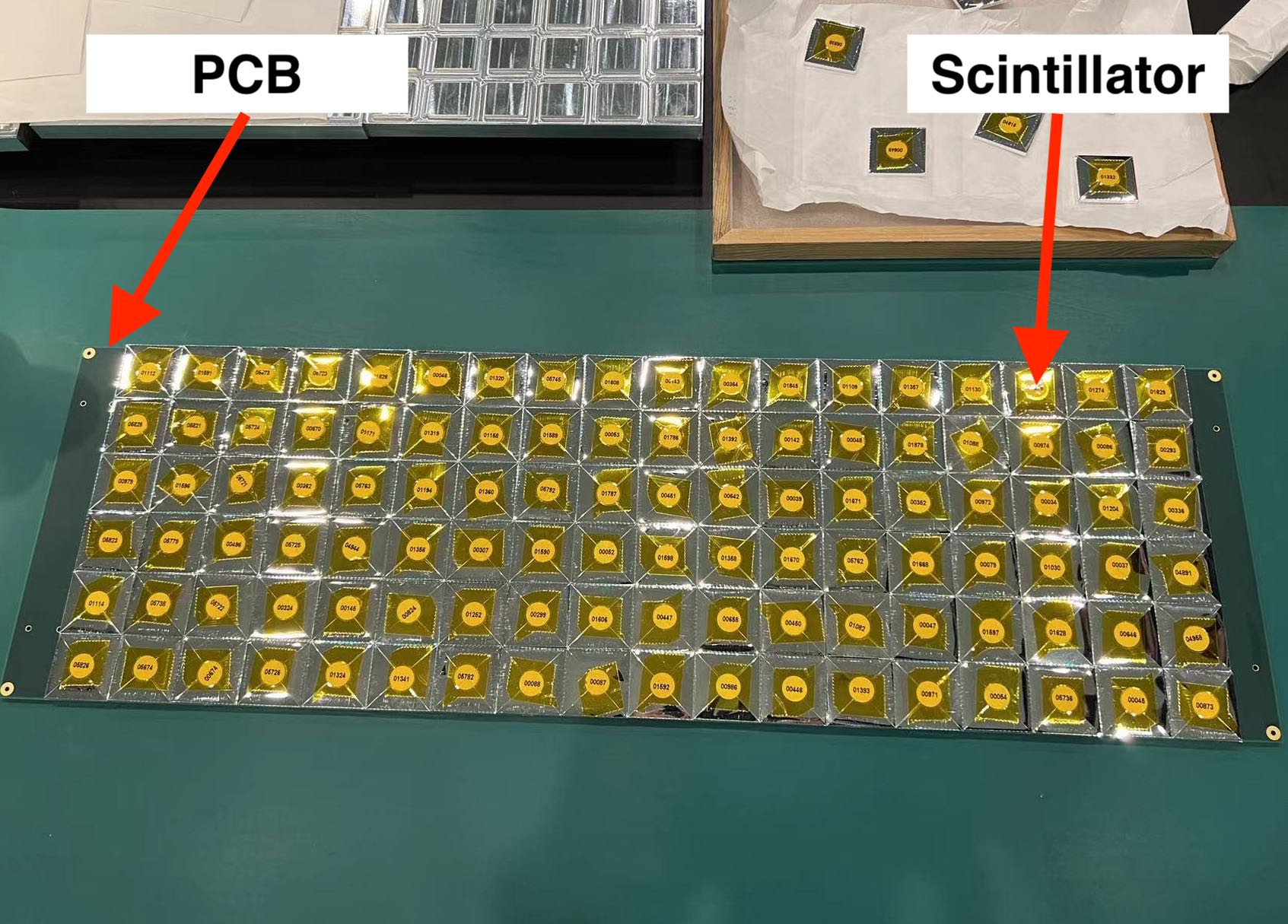}
   
}

\caption{The schematic sandwich structure of AHCAL prototype (a). The array of $4\times4\times0.3\ \mathrm{cm^3}$ scintillator tiles on a PCB (b). One sampling layer contains three PCBs.}
\label{fig:ahcal}
\end{figure}

The detector used in this study is the CEPC Analogue Hadronic Calorimeter (AHCAL) prototype featured in high granularity \cite{li2021optimization, duan2022scintillator, shi2022design}. As depicted in figure \ref{fig:ahcal}, the AHCAL prototype is constructed with 40 sampling layers containing steel cassettes and absorbers. The size of the AHCAL prototype is around $72 \times 72 \times 120\ \mathrm{cm^3}$, with approximately 5 nuclear interaction lengths ($\lambda_{I}$) in the longitudinal direction, a sampling fraction of approximately 2.5\%, and $12,960$ readout channels in total. In a steel cassette, there are scintillator tiles arranged in an $18\times18$ array. Each scintillator tile measures approximately $4\times4\times0.3\ \mathrm{cm^3}$ and is coupled with a silicon photomultiplier (SiPM). The thickness of steel is $2 \ \text{cm/layer}$.

 In this context, 960,000 simulation shower samples of single-primary-particle events, perpendicularly entering the AHCAL, are prepared. These samples are represented by 3D images ($18\times18\times40$ pixels). To be more specific, the Geant4 11.1.1 Toolkit \cite{agostinelli2003geant4} with the $\mathrm{QGSP\_BERT}$ \cite{apostolakis2009progress} physics list was employed to conduct a comprehensive investigation into the response of pions and electrons within calorimeters, aiming to examine the behavior of these particles under similar conditions as the experimental setup. In figure \ref{fig:shower}, we can find that the shower topology of MC samples is close to the shower topology of test beam data, from the perspective of the human eye. At present, test beam data collected at CERN SPS beamline \cite{de1985cern} in 2022 and 2023 are still under calibration and check. Therefore, we only employ MC samples in the following sections. It is still sufficient to effectively compare all PID methods since the same MC samples are employed. As listed in table \ref{tab:data_set}, a wide energy range of 5~GeV--120~GeV has been explored, with a uniform mixture of electron and pion events.

\begin{table}[h]
   
    \caption{ The composition of MC samples.}
    \centering
    \label{tab:data_set}

    \resizebox{0.95\textwidth}{!}{
    \fontsize{35}{40}\selectfont
   
    \begin{tabular}{lcccccccccccccccccc}
    
    \toprule
  
    \multirow{2}{*}{Energy point} & \multicolumn{2}{c}{5 GeV} & \multicolumn{2}{c}{10 GeV} & \multicolumn{2}{c}{30 GeV} & \multicolumn{2}{c}{50 GeV} & \multicolumn{2}{c}{60 GeV} & \multicolumn{2}{c}{80 GeV} & \multicolumn{2}{c}{100 GeV} & \multicolumn{2}{c}{120 GeV} \\ 
    \cmidrule(r){2-3} \cmidrule(r){4-5} \cmidrule(r){6-7} \cmidrule(r){8-9} \cmidrule(r){10-11} \cmidrule(r){12-13} \cmidrule(r){14-15} \cmidrule(r){16-17} 
    
    ~ & $\#$ & Source & $\#$ & Source & $\#$ & Source & $\#$ & Source & $\#$ & Source & $\#$ & Source & $\#$ & Source & $\#$ & Source  \\
    
    Electron & 60k & MC & 60k & MC & 60k & MC & 60k & MC & 60k & MC & 60k & MC & 60k & MC & 60k & MC \\
    
    Pion & 60k & MC & 60k & MC & 60k & MC & 60k & MC & 60k & MC & 60k & MC & 60k & MC & 60k & MC \\

    \bottomrule
    \end{tabular}
    }

\end{table}

In order to counteract the risk of overfitting, these samples are subjected to a randomized split into three distinct sets: the training set, validation set, and the test set in a ratio of 2:1:7. The training set and the validation set are used for building the related BDT and ANN classifiers, while the performance of the constructed BDT and ANN classifiers is evaluated using the test set. Each individual sample undergoes the transformation into input compatible with both BDT and ANN methodologies, ensuring a comprehensive and consistent approach to our analysis.

\section{PID based on BDT}
\label{sec:bdt}

\begin{figure}[htbp]
\centering

\includegraphics[width=0.7\textwidth]{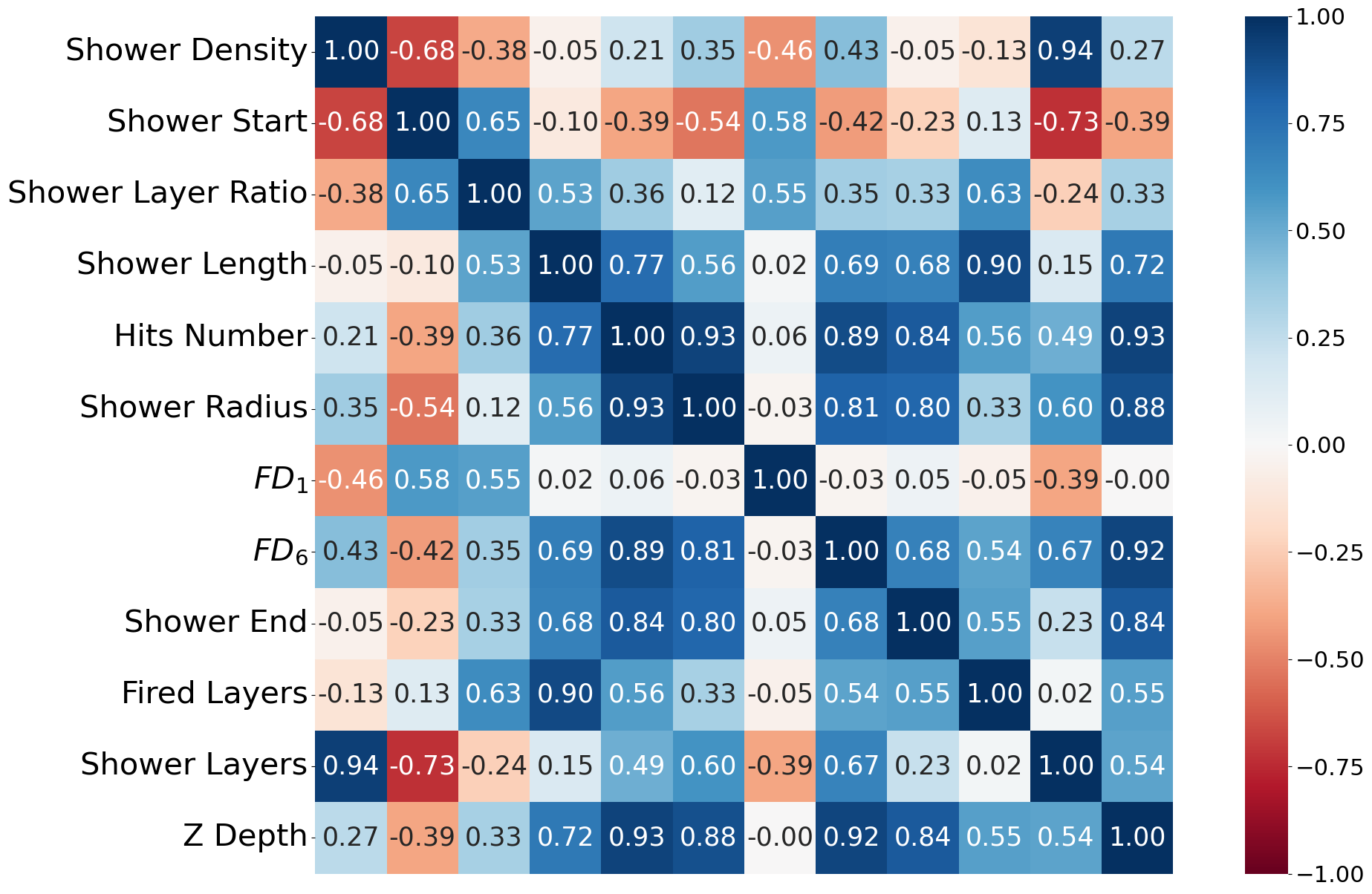}
   
\caption{The correlation matrix of shower topology variables based on MC samples.}
\label{fig:coor}
\end{figure}

We apply Extreme Gradient Boosting (XGBoost), which is known for its high performance\cite{chen2016xgboost}. 12 variables are reconstructed and utilized as inputs since they can encapsulate characteristics of both electromagnetic and hadronic showers (details of these variables are provided in the Appendix~\ref{ap:bdt}). The corresponding correlation matrix is depicted in figure \ref{fig:coor}. In this context, we consider pion events as signals and electron events as backgrounds. Our BDT classifier condenses these 12 reconstructed shower topology variables into BDT pion likelihoods ($L_{\pi}^\mathrm{BDT}$). It is worth highlighting that among these variables, Z Depth and Shower Radius emerge as the two most crucial ones in terms of their separation power, as demonstrated in table \ref{tab:bdt_var_rk_e_mc}.

\begin{table}[h]
    \centering
    \caption{Ranking of separation importance between electromagnetic showers and hadronic showers.}
    \label{tab:bdt_var_rk_e_mc}
    \centering

    \begin{tabular}{l||c}     
    \toprule
    Rank: Variable & Variable weight \\
    \hline
    1: Z depth &\ 0.532 \\
    2: Shower radius &\ 0.186 \\
    3: Shower layers &\ 0.073\\
    4: Fired layers &\ 0.065 \\
    5: Shower density &\ 0.370 \\
    6: Shower start &\ 0.026 \\
    7: Shower layer ratio &\ 0.022 \\
    8: $\mathrm{FD_1}$ &\ 0.018 \\
    9: Hits number &\ 0.013 \\
    10: $\mathrm{FD_6}$ &\ 0.012 \\
    11: Shower end &\ 0.009 \\
    12: Shower length &\ 0.006 \\
    \bottomrule
    \end{tabular}

\end{table}

\newpage

\subsection{BDT method performance}
\label{sec:bdt_perfortmance}

\begin{figure}[htbp]
\centering

\subfigure[]
{   \label{sub:bdt_pi_out_mc}
    
    \centering
    \includegraphics[width=0.45\textwidth]{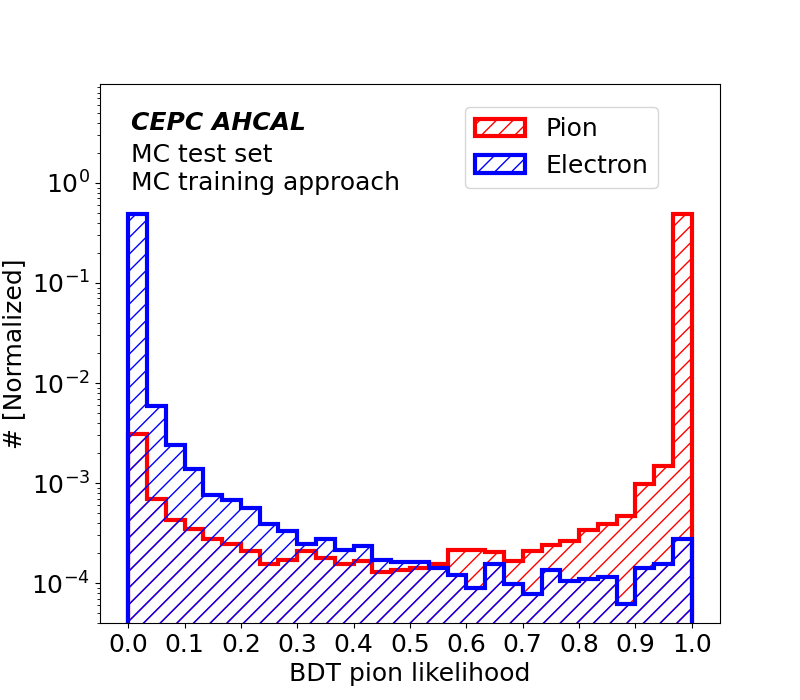}
   
}
\subfigure[]
{   \label{sub:bdt_pi_threshold_mc}
    
    \centering
    \includegraphics[width=0.45\textwidth]{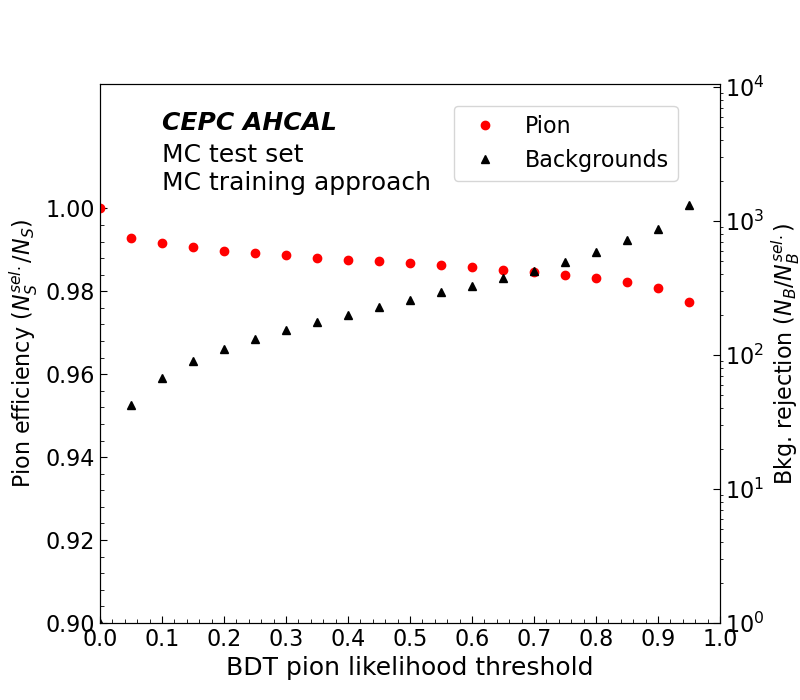}
   
}

\caption{The distribution of the BDT pion likelihood (a). Pion efficiency (in red) and background rejection (in black) as the function of the BDT pion likelihood threshold (b).}
\label{fig:bdt_output}
\end{figure}

 A higher BDT likelihood value indicates a more signal-like event. In figure \ref{sub:bdt_pi_out_mc}, the output BDT likelihoods are presented using the MC test set. The values differ significantly for signal and background. This is confirmed in figure \ref{sub:bdt_pi_threshold_mc}, through two metrics \textemdash\ the signal efficiency ($\varepsilon=N_{S}^{sel.}/{N_{S}}$), representing the proportion of the true signal events selected by our classifier, and the background rejection ($R={N_{B}}/N_{B}^{sel.}$), representing the ratio of the number of all true background events to the number of the background events mistakenly selected as signal events by our classifier (background rejection can also be viewed as the reciprocal of the background efficiency). These two metrics are inversely related. An effective classifier should keep both two metrics robust. 

    
        
 
 In our test set, the number of each type of particle is the same. As reference points in pion identification, to achieve pion signal efficiency ($\varepsilon_{\pi}$) of 99\%, the BDT pion likelihood ($L_{\pi}^\mathrm{BDT}$) should be around 0.2, with corresponding electron background rejection ($R_{e}$) of 105.

\subsection{Dependence of BDT performance on input variables}
\label{subsec:ablation}

An ablation study is conducted to observe the dependence of BDT performance on input variables. Our original BDT classifier has 12 inputs. We first remove variables Shower End, Shower Layers, Fired Layers, and Z Depth, to build BDT with 8 inuts. We further remove variables $\mathrm{FD_1}$ and $\mathrm{FD_6}$, to build BDT with 6 inputs.

\begin{figure}[h]
\centering

\includegraphics[width=0.6\textwidth]{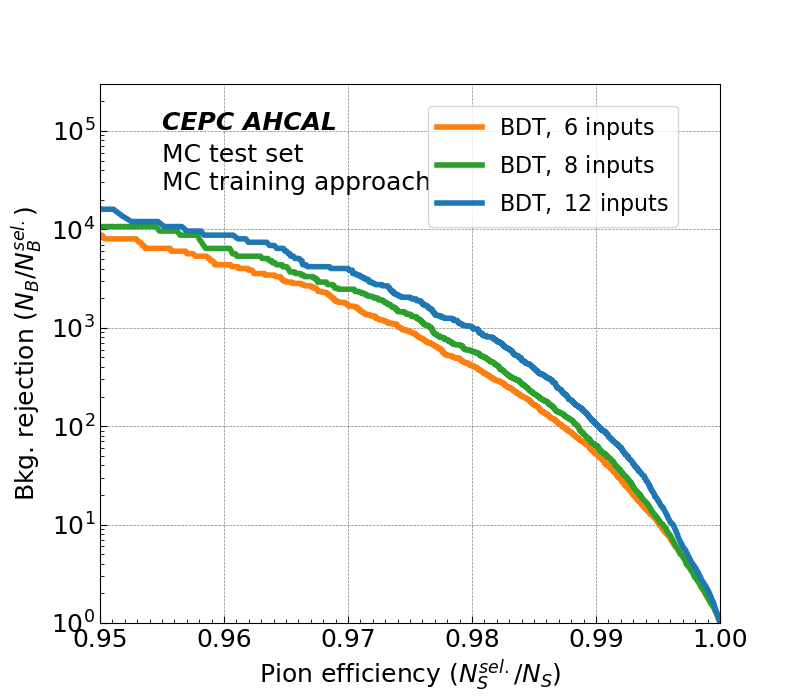}

\caption{The comparison of BDT classifier with different input numbers in terms of background rejection versus signal efficiency in pion identification}
\label{fig:bdt_cp}
\end{figure}

Figure \ref{fig:bdt_cp} illustrates the obvious influence of variable numbers on BDT. When more variables are included, the performance of the BDT classifiers is improved, as the background rejection is higher almost at each signal efficiency point. We assume that the performance of the BDT classifier can be further improved when more meaningful shower topology variables are reconstructed. The further optimization of our BDT classifier is ongoing. In this paper, we only focus on providing a benchmark for our CEPC ResNet classifier. 

\section{PID based on ANN}
\label{sec:ann}

\subsection{Algorithm and architecture}
\label{sec:ann_algo}

An ANN comprises multiple layers of interconnected nodes or ``neurons'' \cite{murtagh1991multilayer, taud2018multilayer}. The input layer receives a tensor. In the context of the CEPC AHCAL, for each event, the energy depositions on the $40\times18\times18$ array of sensitive cells can be directly represented as the input tensor. It is then propagated through the multiple hidden layers with trainable parameters to the output layer, in which intermediate computations and non-linear transformations are made to extract useful features. The output of the ANN is a multi-class tensor $\hat{y}$ with $n$ values, where $n$ is the number of potential primary particle candidates. These values can be normalized using the Softmax function \ref{eq1} \cite{lecun2002efficient}, and interpreted as likelihoods for each candidate. By selecting the candidate with the highest likelihood value, the type of the primary particle is predicted \cite{bishop2006pattern}. 

\begin{equation}
\begin{split}
    &\sigma(\textbf{z})_{i}=\frac{e^{z_i}}{\sum_{j=1}^{n}e^{z_j}}, \\ 
    &\mathrm{for}\ i=1,...,n \ \space \mathrm{and} \ \textbf{z}=(z_1,...,z_n)\in \mathbb{R}^n.
 \label{eq1}
 \end{split}
\end{equation}

\begin{figure}[htbp]
\centering
\subfigure[]
{   
    \label{sub:res18}
    \includegraphics[width=0.9\textwidth]{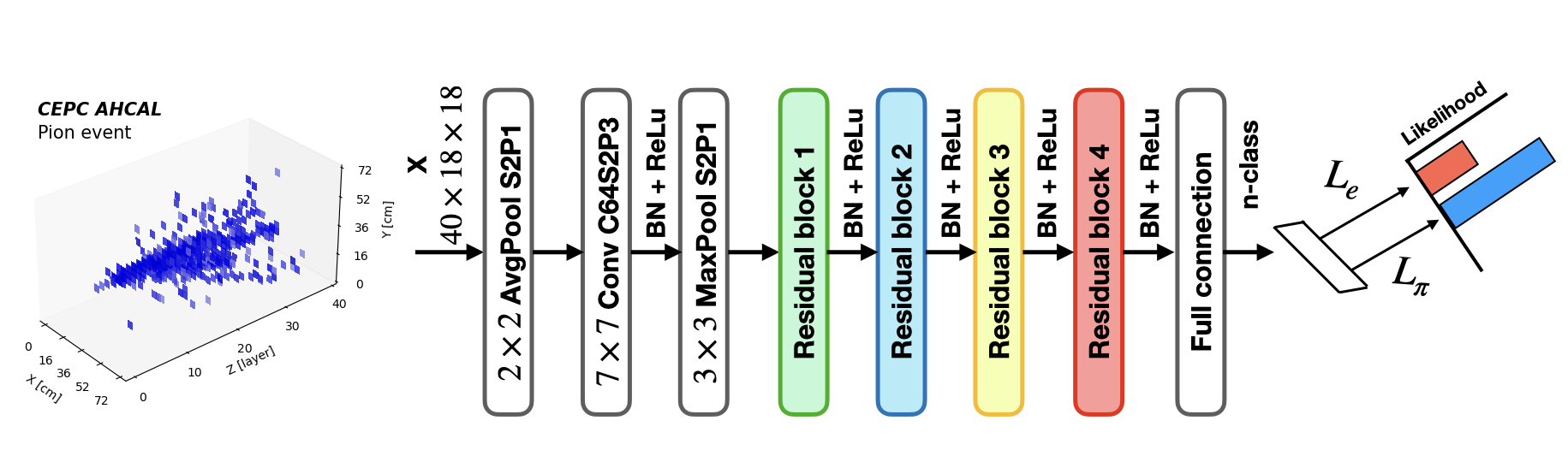}
}

\subfigure[]
{       
    \label{sub:block1}
    \includegraphics[width=0.45\textwidth]{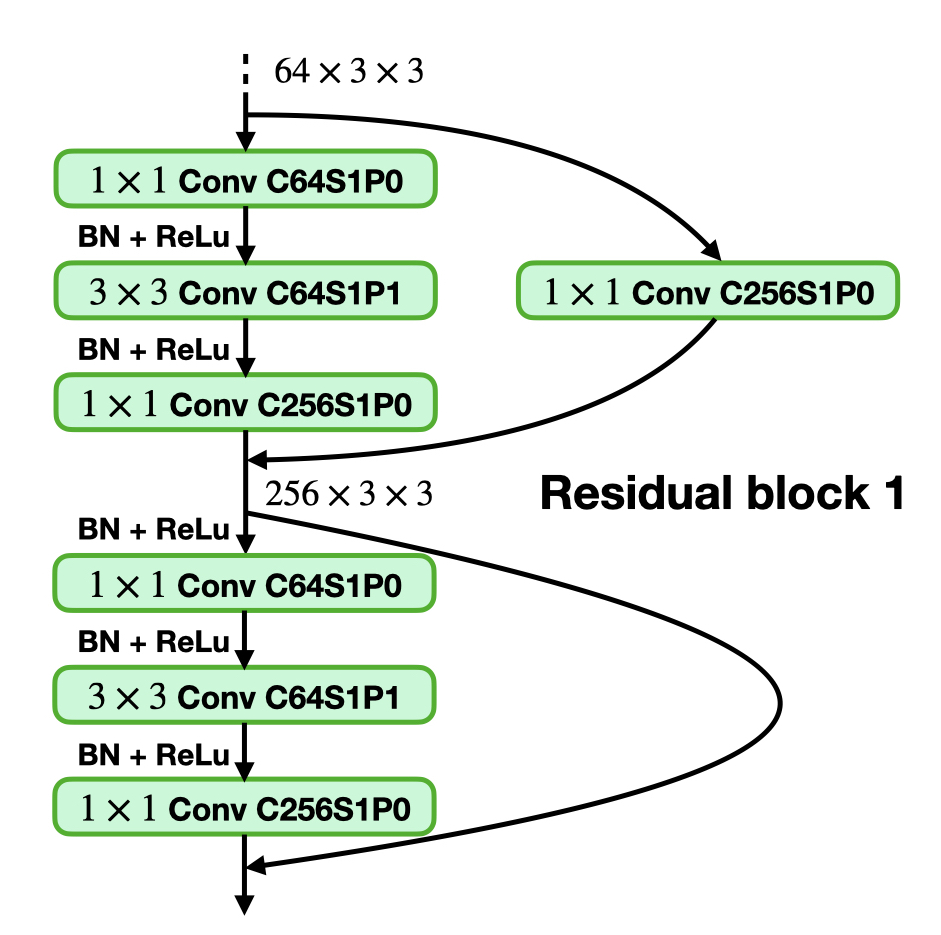}
}
\subfigure[]
{       
    \label{sub:block2}
    \includegraphics[width=0.45\textwidth]{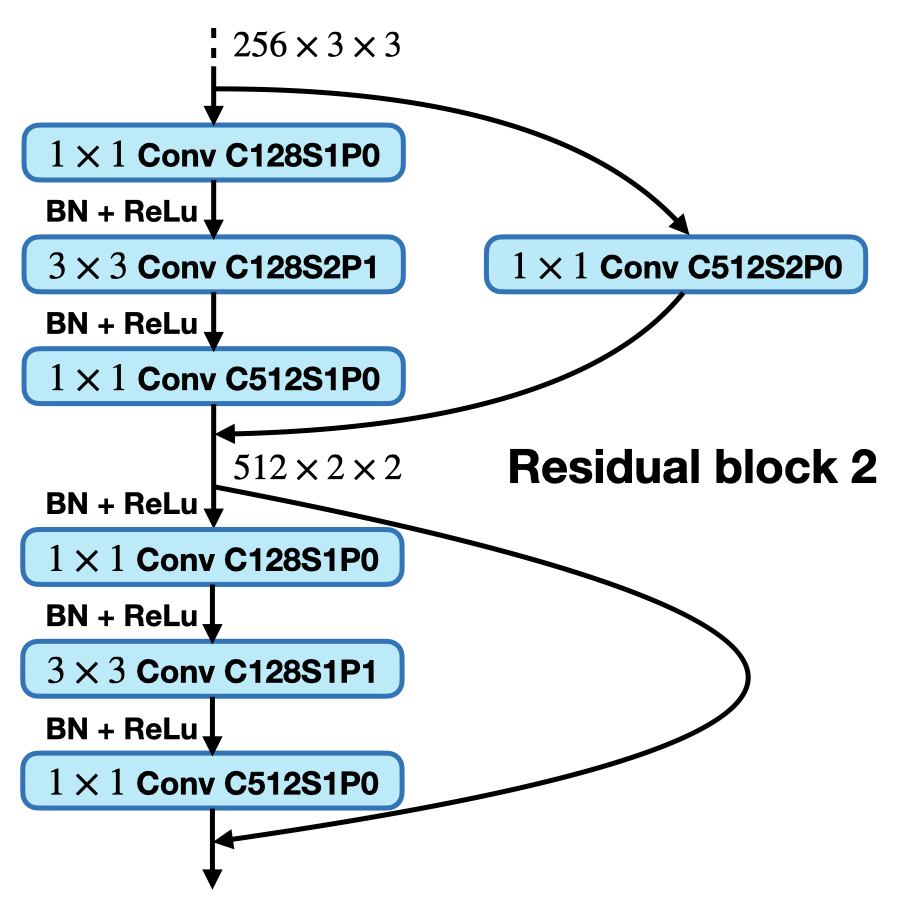}
}

\subfigure[]
{       
    \label{sub:block3}
    \includegraphics[width=0.45\textwidth]{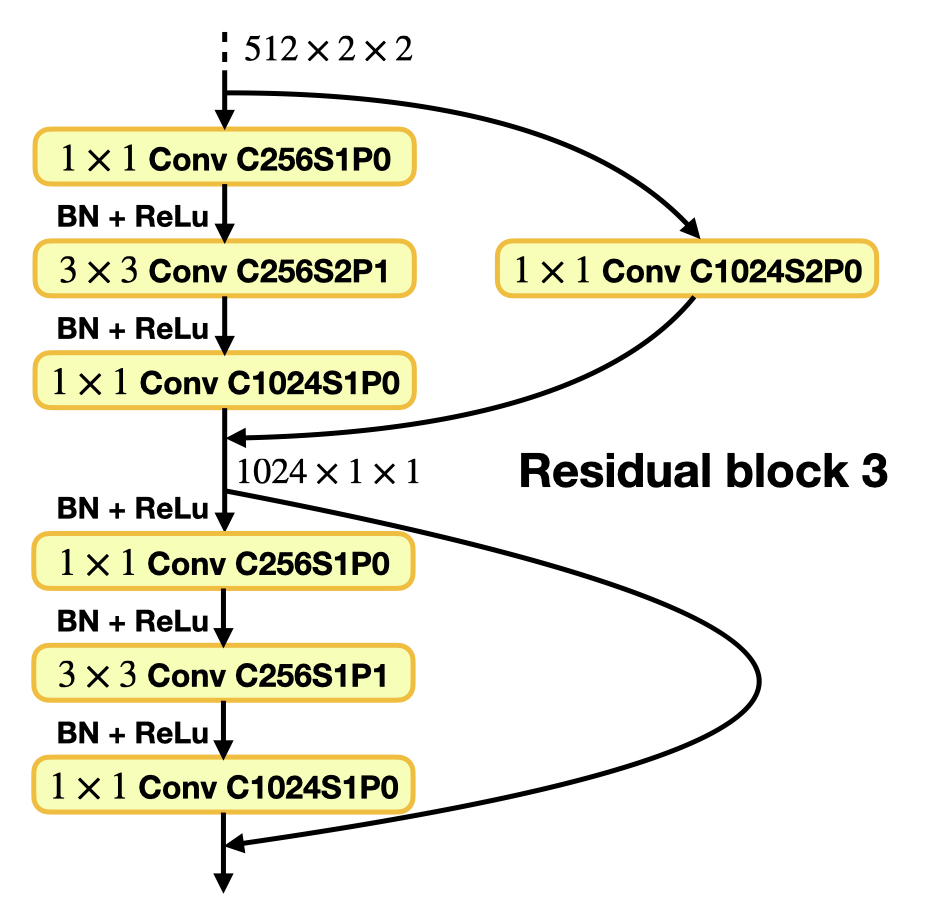}
}
\subfigure[]
{       
    \label{sub:block4}
    \includegraphics[width=0.45\textwidth]{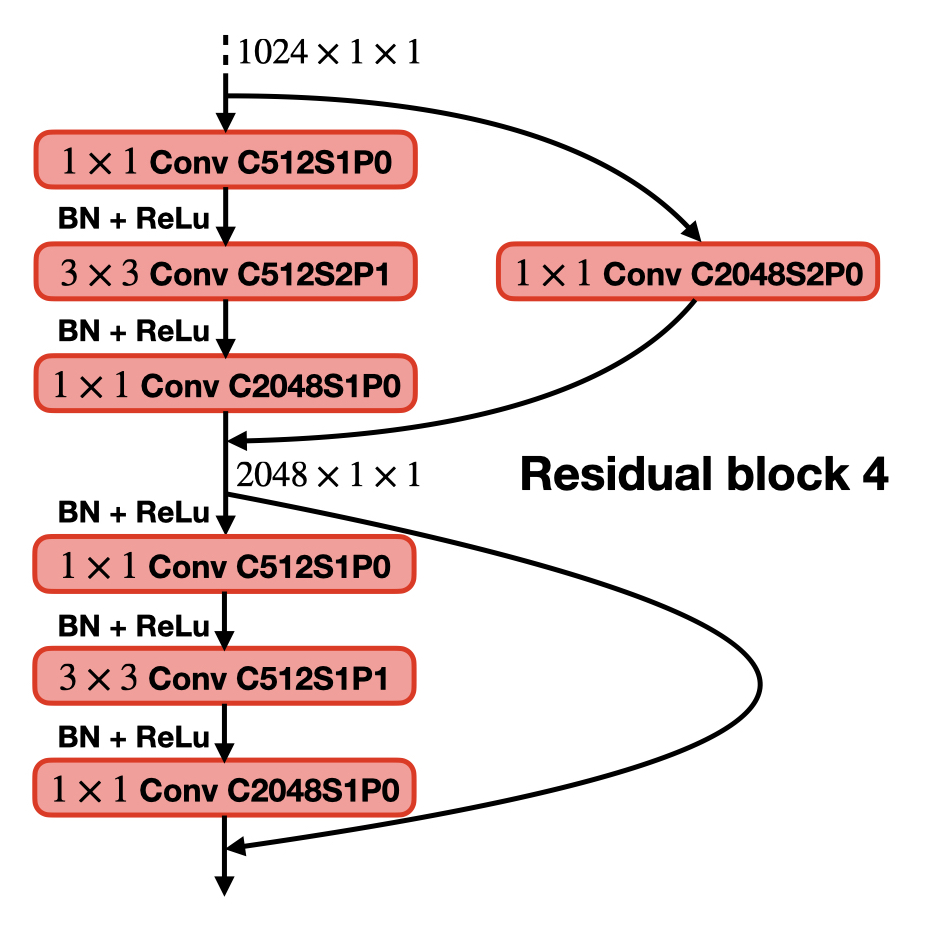}
}

\caption{The schematic net architecture of CEPC ResNet PID. The output is a normalized tensor. The\ ``Conv'' indicates a 2D convolutional layer, and the\ ``C'' specifies the number of output channels. The\ ``S'' denotes the stride, which determines the step size for the convolution operation as the kernel moves over the input. The\ ``P'' represents the padding applied to the input. The\ ``Maxpool'' represents a max-pooling layer. }
\label{fig:res18}
\end{figure}

The structure and complexity of the network can be customized for different tasks \cite{khan2019analysis,wang2017generative,tenney2019bert}. We take the architecture of the ResNet with higher classification accuracies \cite{he2016deep} as demonstrated in figure \ref{fig:res18}. Its generalization ability on unseen data has also been theoretically guaranteed \cite{he2020resnet}. The key idea behind ResNet is the use of residual blocks, which allow the network to learn residual mappings instead of trying to learn the direct mappings between the block's input and output. A residual block consists of convolutional layers to extract features from the input and a shortcut connection, also known as a skip connection, that bypasses one or more layers. By propagating the identity mapping through the shortcut connection, ResNet enables the network to learn the residual mapping, capturing the difference between the input and output. In order to adapt it to 40 sampling layers of the AHCAL prototype, the input channel number of the first convolution layer has been set to 40. Moreover, as shown in figures \ref{sub:block1}, \ref{sub:block2}, \ref{sub:block3}, and \ref{sub:block4}, architectures in residual blocks have also been adjusted as more convolutional layers are included.

The trainable parameters are automatically adjusted during the training process to improve the ability of ResNet to make accurate predictions. This adjustment is achieved by minimizing a loss function that quantifies the discrepancy between the network's predicted outputs and the ground truth labels. In our PID method, we employ the categorical cross-entropy loss function \cite{ketkar2017deep}, which is commonly used for image classification tasks and also suits our PID framework. The net's prediction $\hat{y}_i$ serves as the input to the categorical cross-entropy loss function, which becomes a function of the network's trainable parameters $\theta$. This relationship is expressed in equation \ref{eq3}. To optimize the values of $\theta$ and minimize the loss, we employ the Stochastic Gradient Descent (SGD) algorithm \cite{amari1993backpropagation, boyd2004convex, bottou2012stochastic}. This process is outlined in Algorithm \ref{alg1}. Through this optimization process, the ResNet gradually learns to make more accurate predictions, improving its performance in particle identification tasks.

\begin{equation}
\mathrm{Loss}=\mathrm{Loss}\left({y}_{i},\hat{{y_i}}(\theta)\right)=-\sum_{i=1}^{n}y_{i}\log{\hat{y_i}(\theta)} 
\label{eq3}
\end{equation}

\begin{algorithm}[H]
\caption{Artifical Neural Networks.}\label{alg1}
\begin{algorithmic}[1]
\Require{The batch size $m$, the epoch number $n$, initial learning rate $lr$, initial net parameters $\theta_{0}$.}

\State{Assign corresponding label $y$ to data $x$.}
\For{$t=1, \cdots ,k$ iteration steps }
\For{$i=1, \cdots ,m$}

\State $\hat{y} \gets \mathrm{Net} (x,\theta)$
\State $\mathrm{Loss}({y}_{i},\hat{{y_i}})_{\theta}^{(i)} \gets (-\mathrm{log}(\hat{{y_i}}))$
\EndFor
\State $\theta \gets \mathrm{SGD}\left(\bigtriangledown_{\theta} \frac{1}{m}\sum_{i=1}^{m}\mathrm{Loss}_{\theta}^{(i)}\right)$
\EndFor

\end{algorithmic}
\end{algorithm}

To train an ANN, including ResNet, the total iteration steps $k$ equals $N\times{n}/m$, where $N$ is the number of events for training, $m$ is the batch size, and $n$ is the epoch number. The neural network is executed on an NVIDIA V100 NVLink GPU, utilizing the NVIDIA CUDA platform. The code is scripted in Python and employs the PyTorch deep learning library \cite{paszke2019pytorch}. Due to resource constraints, conducting an exhaustive exploration of hyper-parameters was not feasible. However, we still tested hundreds of hyper-parameter combinations, repeated the training 10 times with different initial weights to mitigate the impact of fluctuations resulting from different initializations, and finally selected the one that yielded the best performance (highest classification accuracy). We decided the batch size $m$ to be 64, the epoch number $n$ to be 200, and the initial learning rate $lr$ to be $1\times10^{-4}$.

\subsection{Impacts of the Residual Connections}

In our experiment, the efficacy of the Residual Connections is first validated by comparing the error ($\frac{N_{misclassified}}{N_{total}}\times$100\%) during the iteration with and without the Residual Connections shortcut across various combinations of the batch size (64,~ 128,~ 256~) and the learning rate ($1\times10^{-5}$, $1\times10^{-4}$, $1\times10^{-3}$) while maintaining 200 epochs (figures of the training error and the validation error versus iteration are provided in the Appendix \ref{ap:error}). Notably, the number of iterations per epoch varies, with 3000 iterations per epoch for a batch size of 64, 1500 iterations per epoch for a batch size of 128, and 750 iterations for a batch size of 256. When the network is operated without the shortcut, our ResNet classifier will become non-residual.

It is observed that the existence of the shortcut connection leads to a faster decrease in both training and validation errors during the initial few epochs. Moreover, the existence of the shortcut connection mitigates the fluctuation of the validation error during the iteration, which is observed across all combinations of the batch size and the learning rate, consistent with findings in the reference paper \cite{he2016deep}. Table \ref{tab:shortcut} lists the validation error recalculated after 200 epochs of iterations using the validation set with various combinations of the batch size and the learning rate. Under the optimized combination of the learning rate and the batch size, the validation error is at its lowest, standing at 0.35\%. Introducing the shortcut connection results in a reduction of the validation error by a factor of 2-3 in most cases of our experiments. These results suggest that the existence of the shortcut can enhance the robustness of the network and should be employed in our model.

\begin{table}[h]
    \renewcommand{\arraystretch}{1}
    \caption{The validation error (\%) was recalculated after 200 epochs' iterations using the validation set. Keeping the learning rate and batch size constant, the only distinction lies in the presence or absence of the shortcut. The training procedures are illustrated in figure \ref{fig:error_shortcut}. }
    \centering
    \label{tab:shortcut}
       
    \begin{tabular}{l||cc}
    \toprule
    & W/ shortcut& W/O shortcut\\
    \hline
    $lr$ = $1\times10^{-5}$, batch = 64  & 0.441 & 0.716 \\
    $lr$ = $1\times10^{-5}$, batch = 128 & 0.497 & 2.445 \\
    $lr$ = $1\times10^{-5}$, batch = 256 & 0.621 & 5.789 \\
    $lr$ = $1\times10^{-4}$, batch = 64  & \textbf{0.350} & 2.009 \\
    $lr$ = $1\times10^{-4}$, batch = 128 & 0.407 & 1.220 \\
    $lr$ = $1\times10^{-4}$, batch = 256 & 0.397 & 1.385 \\
    $lr$ = $1\times10^{-3}$, batch = 64  & 0.371 & 0.615 \\
    $lr$ = $1\times10^{-3}$, batch = 128 & 0.378 & 1.015 \\
    $lr$ = $1\times10^{-3}$, batch = 256 & 0.409 & 1.156 \\

    \bottomrule
    \end{tabular}
    
\end{table}

\subsection{ResNet performance}
\label{subsec:ann_perfortmance}

\begin{figure}[htbp]
\centering

\subfigure[]
{   \label{sub:ann_pi_mc}    
    \includegraphics[width=0.45\textwidth]{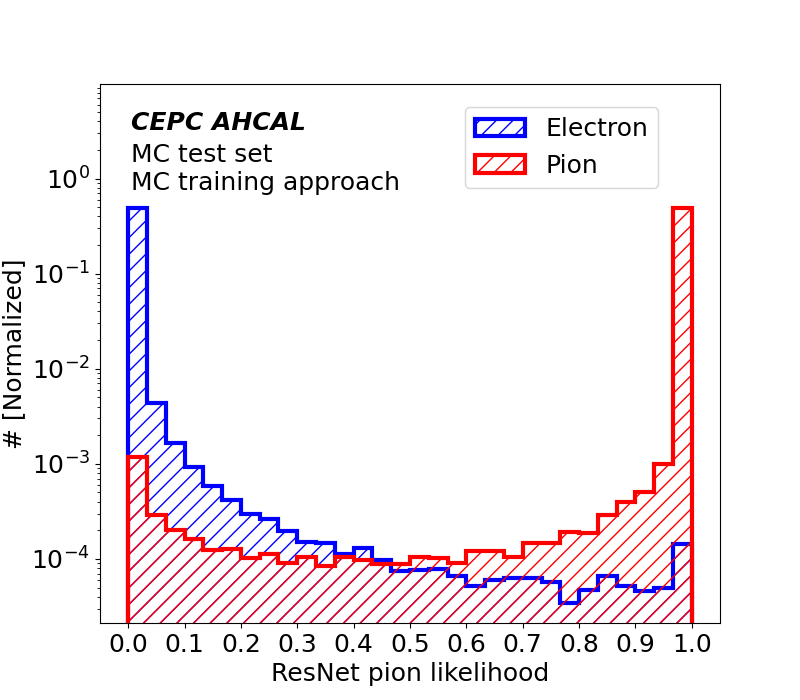}   
}
\subfigure[]{ 
    \label{sub:ann_pi_mc_s_b_threshold}
    \includegraphics[width=0.45\textwidth]{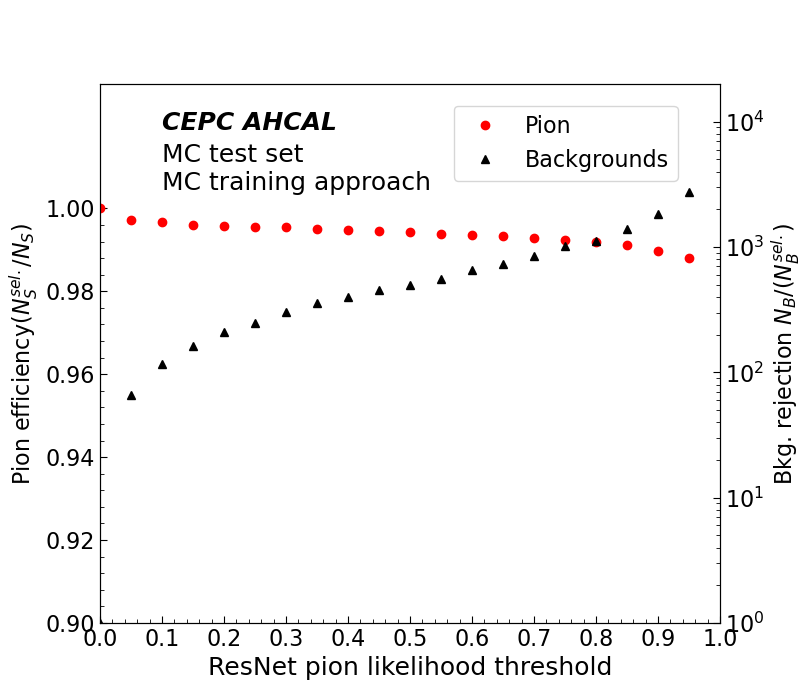}
}

\caption{The distribution of the ResNet pion likelihood (a). Pion efficiency (in red) and background rejection (in black) as the function of the ResNet pion likelihood threshold (b).}
\label{fig:ann_output_mc}
\end{figure}

As the same, electron events and pion events in the test set occupy the same proportion. Figure~\ref{sub:ann_pi_mc} shows the pion likelihood distribution. It can be observed that a significant separation power has also been achieved. For example, in figure \ref{sub:ann_pi_mc}, pion events tend to have higher values in the pion likelihood compared to electron events, indicating that the ResNet is capable of distinguishing pions from electrons. This observation is further supported by figure \ref{sub:ann_pi_mc_s_b_threshold}. At the same target pion efficiency ($\varepsilon_{\pi}$) of 99\%, the ResNet achieves an electron background rejection ($R_{e}$) of approximately 2000, where the ResNet pion likelihood ($L_\pi^\mathrm{ResNet}$) is around 0.9.

\begin{figure}[h]
\centering

\includegraphics[width=0.6\textwidth]{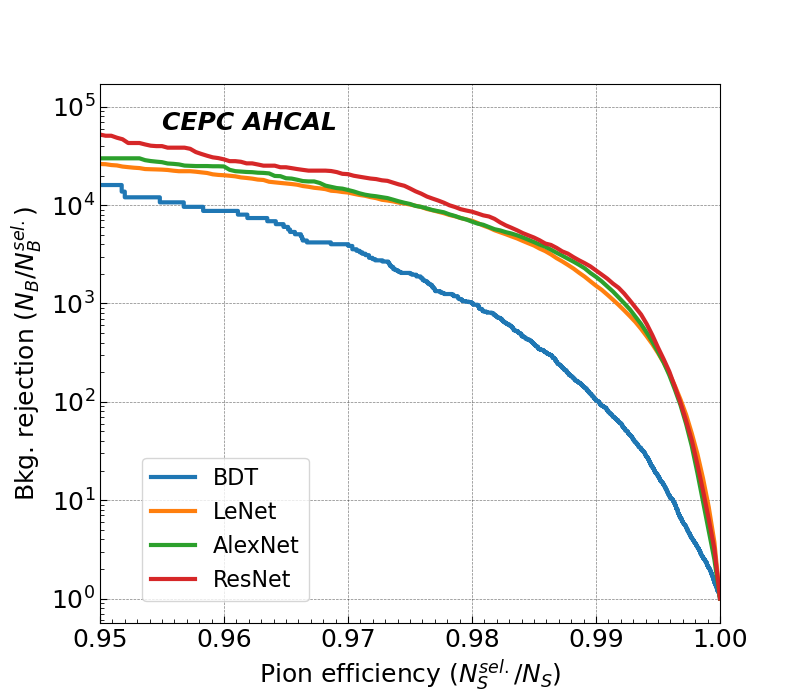}

\caption{The comparison of background rejection versus signal efficiency for BDT, LeNet, AlexNet, and ResNet in identifying pions against electron backgrounds, with results computed from the test set.}
\label{fig:bkg_ratio_mc_compare}
\end{figure}

Moreover, our ResNet classifier is compared with the aforementioned BDT classifier  (utilizing 12 input variables), as well as with LeNet\cite{lecun1998gradient}, one of the earliest CNN, and AlexNet\cite{krizhevsky2012imagenet}, which achieved remarkable performance on the ImageNet Large Scale Visual Recognition Challenge (ILSVRC)\cite{russakovsky2015imagenet} in the past. It is noteworthy that all neural network models undergo an identical hyper-parameter optimization process, and the training is conducted 10 times with varying randomly initialized weights to evaluate stability. The combination of hyper-parameters (batch size, learning rate) is (32, $1\times10^{-3}$) for AlexNet and (256, $1\times10^{-5}$) for LeNet. Figure \ref{fig:bkg_ratio_mc_compare} presents the comparison in terms of electron background rejection at the same pion signal efficiency and table \ref{tab:ann_pid_performance} summarizes the corresponding key metrics with physics insights including electron rejection ($R_e$) at a certain pion efficiency ($\varepsilon_{\pi}$) of 97\%, 98\%, 99\%, and 99.5\%, and the classification accuracy. We observe that the ResNet classifier tailored by our team demonstrates superior performance. For instance, at $\varepsilon_{\pi}$ of 99\%, The ResNet classifier markedly enhances $R_{e}$ by a factor of 20, increasing it from approximately 100 to about 2000. Furthermore, when $\varepsilon_{\pi}$ equals 97\%, ResNet's $R_e$ capability surpasses that of AlexNet and LeNet by roughly 1.5 times, and it even outperforms BDT by an impressive factor of 5.


   


\begin{table}[h]
    \renewcommand{\arraystretch}{1}
    \caption{PID performance of BDT, AlexNet, LeNet, and ResNet on pion identification, evaluated using the test set. The training of BDT is performed only once since the uncertainty of BDT due to random weight initialization is expected to be fairly small. The training of other models is repeated 10 times using different randomly initialized weights. The table shows the mean value and the standard error of electron background rejection ($R_e$) and classification accuracy to assess the stability to random weight initialization. The uncertainty of the classification accuracy of LeNet is also found to be negligible.}
    \centering
    \label{tab:ann_pid_performance}
       
    \begin{tabular}{lccccc}
    \toprule
    & $R_{e}^{\varepsilon=97\%}$ & $R_{e}^{\varepsilon=98\%}$ & $R_{e}^{\varepsilon=99\%}$ & $R_{e}^{\varepsilon=99.5\%}$ & Accuracy [\%]\\
    \hline
     BDT & 4003 & 1011 & 105 & 18 & 99.15\\
    
    AlexNet  & 14260 $\pm$ 223  &6791 $\pm$ 73 & 1864 $\pm$ 36 & 323 $\pm$ 7 & 99.45 $\pm$ 0.04\\ 
    
    LeNet  & 13427 $\pm$ 163  & 6933 $\pm$ 49 & 1514 $\pm$ 19 & 314 $\pm$ 4 & 99.43\\
    
    ResNet  &\textbf{23576 $\pm$ 202}  & \textbf{8547 $\pm$ 85} & \textbf{2165 $\pm$ 45} & \textbf{356 $\pm$ 6} & \textbf{99.61 $\pm$ 0.02}\\
    \bottomrule
    \end{tabular}
      
\end{table}

\subsection{Prospects in point cloud scenarios}
\label{sec:point_clouds}

Our ResNet PID model currently relies on detectors with regular structures to reconstruct image-based inputs, though the AHCAL prototype exhibited normal structures, making it a natural choice to employ 2D-convolution-based deep learning methods. However, it is important to note that the geometry of detectors is not universally confined to cubic structures with fixed tile sizes. For example, detectors utilized in collider experiments, both existing and future ones\cite{collaboration2008cms, aad2008atlas, huffman2014plans, bunin2023upgrade}, often adopt a barrel-like configuration with variable tile sizes, to maximize the chances of detecting particles produced in the collision. The upgraded design of the AHCAL in the future would probably not be limited to fixed tile size as well. Therefore, this might pose challenges to the application of our algorithm. 

3D shapes could also be represented in the point clouds, characterized by their permutation invariance and the flexibility to incorporate diverse features for each point. This approach ensures that the representation of showers is not constrained by the geometry of detectors, thereby showcasing its versatility as an input method. Here, we propose a tailored solution to point cloud scenarios based on the current ResNet model, called the Dynamic Graph Residual Networks (DGRes).

\begin{figure}[h]
\centering

\includegraphics[width=0.95\textwidth]{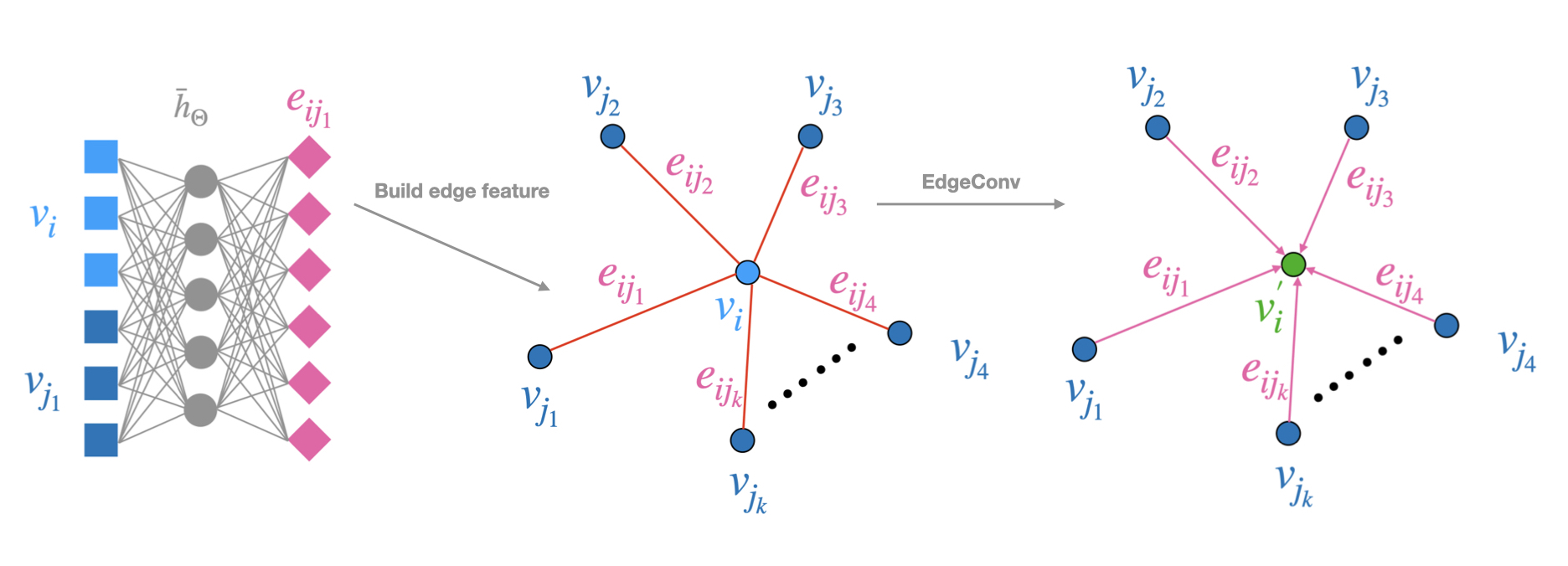}

\caption{The schematic plot illustrates the functionality of the EdgeConv block. In this example, the edge feature $e_{ij_{1}}$ is computed from a pair of vertices, $v_i$ and $v_{j_{1}}$, by instantiating $\bar{h}_\Theta$ with a fully connected layer to map ($v_i$, $v_{j_{1}}$-$v_i$) into $e_{ij_{1}}$. The output of EdgeConv, denoted as $v_i'$, is calculated by aggregating the edge features associated with k nearest vertices.}
\label{fig:dgcnn}
\end{figure}

 Traditional approaches to processing point clouds typically rely on handcrafted features to capture their geometric properties \cite{rusu2009fast, lu2014recognizing}. Recently, advancements in deep point cloud processing methods by leveraging deep neural networks have been made. These data-driven methods excel in learning features and have demonstrated superior performance compared to traditional techniques across a spectrum of tasks \cite{chang2015shapenet, zhang2019review}. DGRes is inspired by the Dynamic Graph Convolutional Neural Networks (DGCNN)\cite{wang2019dynamic}, which introduces graph convolution, namely the EdgeConv, to explore structural information between points. Figure\ref{fig:dgcnn} illustrates the fundamental concept of EdgeConv, which involves treating points themselves as vertices ($v_i$), with edges established as connections between dynamically changed $k$ nearest neighboring vertices, determined by the k-nearest neighbor (k-NN) algorithm \cite{peterson2009k}. The same formulation of EdgeConv from the referenced paper \cite{wang2019dynamic} is adopted in our model:

\begin{equation}
\mathbf{v}_i^{\prime}=\underset{j:(i,j)\in\mathcal{E}}{\operatorname*{\mathrm{max}}}e_{ij}=\underset{j:(i,j)\in\mathcal{E}}{\operatorname*{\mathrm{max}}}{\bar{h}_\Theta(\mathbf{v}_i,\mathbf{v}_j-\mathbf{v}_i) }.
\label{eq:edgconv}
\end{equation}

\begin{figure}[h]
\centering

\includegraphics[width=0.85\textwidth]{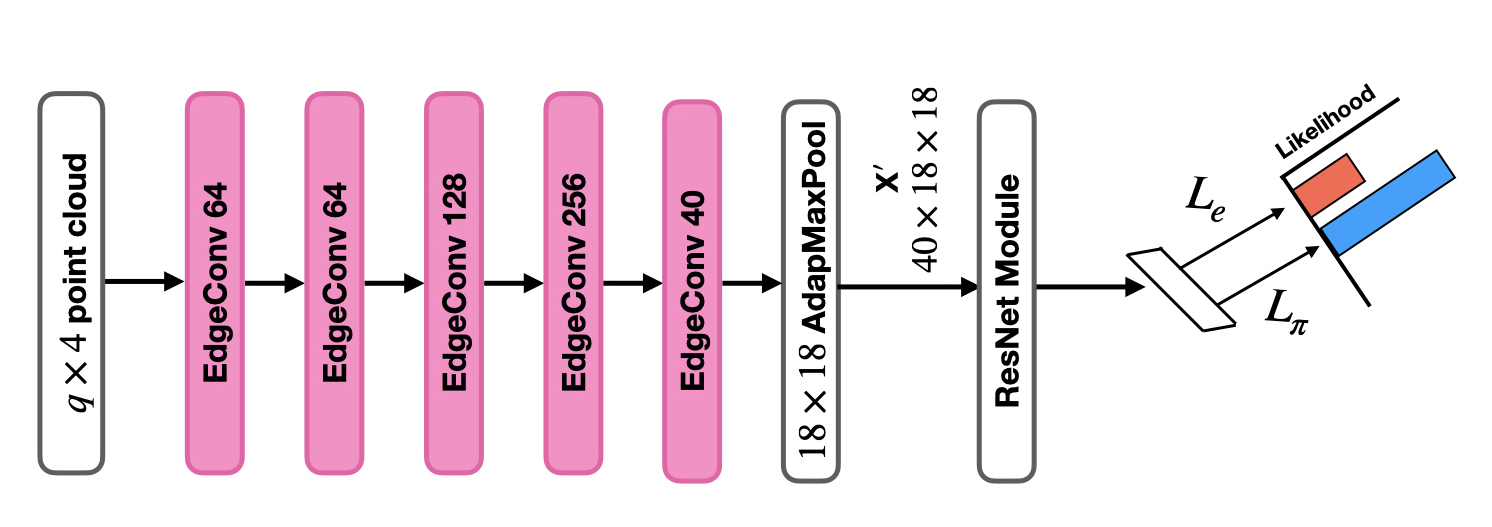}

\caption{The architecture of DGRes. Conversion layers based on EdgeConv take point-cloud-based data as input and map them into tensors with 3 dimensions (shape: $40\times18\times18$), which could then be processed by the ResNet module illustrated in figure \ref{fig:res18}.}
\label{fig:dgres}
\end{figure}

The $\mathbf{V}=\{\mathbf{v}_1,\ldots,\mathbf{v}_q\}\subseteq\mathbb{R}^F$ denotes $q$ vertices in the cloud with features, $e_{ij}$ is the edge feature, and $\bar{\boldsymbol{h}}_{\Theta}$ is realized as a multilayer perceptron (MLP) with shared parameters across all edges. Consequently, the architecture of the DGRes (shown in figure \ref{fig:dgres}) can be conceptually delineated into two parts. The initial part comprises conversion layers based on EdgeConv, to facilitate the transformation of point-cloud-based (graph-based) data into image-based data within a reconstructed three-dimensional space. Subsequently, the latter part adopts the ResNet architecture to extract features and classify events, which is discussed in Subsection \ref{subsec:ann_perfortmance}.

A comparative analysis is conducted involving our DGRes and two baseline algorithms: DGCNN and GravNet. DGCNN has been already introduced above. The GravNet employs distance-weighted graph networks, originally proposed for tasks such as cluster reconstruction and particle identification \cite{qasim2019learning}. The integration of GravNet has resulted in the formulation of an end-to-end reconstruction algorithm tailored for the single-shot calorimetric reconstruction of approximately 1000 particles. This is particularly crucial in high-luminosity conditions with 200 pileup events \cite{qasim2022end}. 

All models undergo training and testing using the identical set of feature vectors pre-reconstructed from the dataset introduced in Section \ref{sec:dataset}. Feature vectors $v_{i}=~(x_i, y_i, z_i, e_i)$ represent the coordinates of hits and the corresponding deposited energy. Training is consistently conducted over 200 epochs. To accommodate variations in the Hits Number for each event, we make $q$, the number of vertices, as a hyper-parameter during training to facilitate efficient mini-batch training. When the Hits Number exceeds $q$, we randomly select $q$ different hits from all hits to form vertices to create a sub-graph in each iteration. Conversely, if the Hits Number is smaller than $q$, features of the empty vertices are padded with zeros. Given that electromagnetic showers and hadronic showers exhibit distinct local structures, the number of nearest neighboring vertices, $k$, can be interpreted as a local connection scale for hits in showers. The effectiveness of models in capturing local structures is influenced by the specific value of $k$, making it another hyper-parameter to be optimized. 

\begin{figure}[h]
\centering

\includegraphics[width=0.6\textwidth]{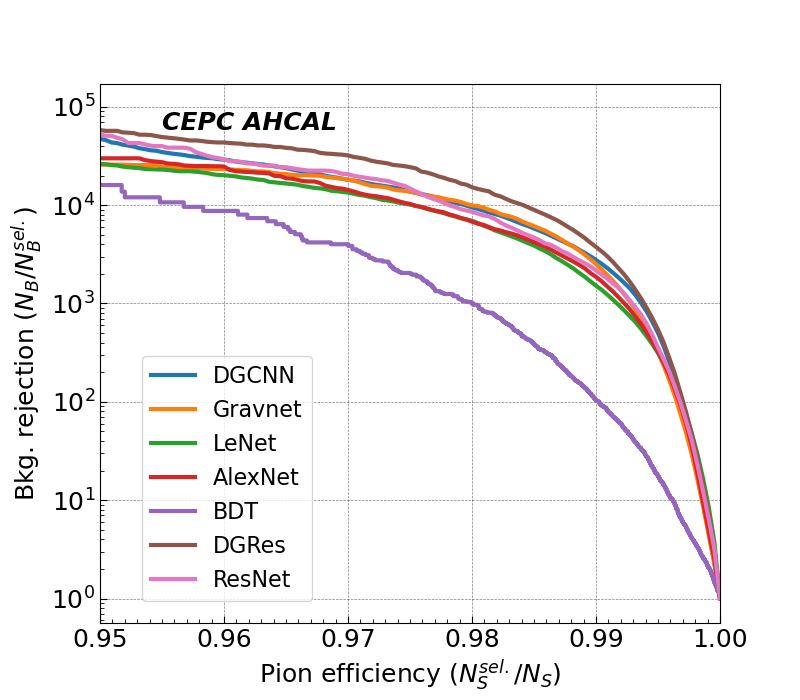}

\caption{The comparison of background rejection versus signal efficiency for DGCNN, GravNet, and DGRes, together with BDT, LeNet, AlexNet, and ResNet, in identifying pions against electron backgrounds, with results computed from the test set.}

\label{fig:point_cloud}
\end{figure}

For DGCNN, we utilize an existing PyTorch implementation. In the case of GravNet, we re-implement the network architecture originally for the reconstruction task, as outlined in the paper \cite{qasim2022end}, into PyTorch for classification. The combination of parameters (batch size, learning rate, number of vertices, number of nearest neighboring vertices) is re-optimized, resulting in (32, 0.01, 256, 10) for DGRes, (128, 0.1, 128, 20) for DGCNN, and (32, $1\times10^{-3}$, 256, 10) for GravNet. The performance evaluation utilizes identical metrics with physics insights outlined in Subsection \ref{subsec:ann_perfortmance}. Figure \ref{fig:point_cloud} depicts the electron background rejection versus the pion signal efficiency of DGRes, DGCNN, and GravNet. Moreover, this comparison encompasses AlexNet, LeNet, BDT, and ResNet, providing a comprehensive overview of all models, with corresponding results summarized in table \ref{tab:ann_pid_performance_pc}.

Notably, the DGRes model demonstrates superior performance across all metrics. Specifically, DGRes achieves a background rejection of 3758, surpassing GravNet and DGCNN by approximately a factor of 1.5 when the signal efficiency ($\varepsilon_{\pi}$) is set at 99\%. This superior performance can be attributed to the utilization of ResNet with deeper networks following several EdgeConv blocks in our DGRes model, which expands the parameter space, promising enhanced performance. Additionally, the effectiveness of EdgeConv blocks in capturing features and transforming point cloud data into image-based data is evident from the improved performance of DGRes compared to purely image-based ResNet, such as improving the electron background rejection by a factor of 1.7 at signal efficiency of 99\%. Consequently, these findings indicate promising prospects for the application of our PID method, even in scenarios where tile sizes vary.

\begin{table}[h]
    \renewcommand{\arraystretch}{1}
    \caption{PID performance of DGRes is compared to GravNet and DGRes on pion identification, evaluated using the test set. The training of each model is repeated 10 times using different randomly initialized weights. The table shows the mean value and the standard error of electron background rejection ($R_e$) and classification accuracy to assess the stability to random weight initialization. The uncertainty of classification accuracy of DGCNN and DGRes is found to be negligible. The performance of BDT, AlexNet, LeNet, and ResNet listed in table \ref{tab:ann_pid_performance} is also provided.}
    \centering
    \label{tab:ann_pid_performance_pc}
       
    \begin{tabular}{lccccc}
    \toprule
    & $R_{e}^{\varepsilon=97\%}$ & $R_{e}^{\varepsilon=98\%}$ & $R_{e}^{\varepsilon=99\%}$ & $R_{e}^{\varepsilon=99.5\%}$ & Accuracy [\%]\\
    \hline
   
    DGCNN  & 18085 $\pm$ 289 & 9498 $\pm$ 102 & 2772 $\pm$ 46 & 501$\pm$ 6 & 99.65\\

    GravNet & 18127 $\pm$ 227  & 9942 $\pm$ 88 & 2516 $\pm$ 34 & 321 $\pm$ 4 & 99.47 $\pm$ 0.08\\
    
    DGRes &  \textbf{32011 $\pm$ 204} &  \textbf{15193 $\pm$ 116} &  \textbf{3758 $\pm$ 20} &  \textbf{543 $\pm$ 6} & \textbf{99.71}\\
    \hline
    BDT & 4003 & 1011 & 105 & 18 & 99.15\\
    
    AlexNet  & 14260 $\pm$ 223  &6791 $\pm$ 73 & 1864 $\pm$ 36 & 323 $\pm$ 7 & 99.45 $\pm$ 0.04\\ 
    
    LeNet  & 13427 $\pm$ 163  & 6933 $\pm$ 49 & 1514 $\pm$ 19 & 314 $\pm$ 4 & 99.43\\
    
    ResNet  & 23576 $\pm$ 202  & 8547 $\pm$ 85 & 2165 $\pm$ 45 & 356 $\pm$ 6 & 99.61 $\pm$ 0.02\\
    \bottomrule
    \end{tabular}
      
\end{table}

\newpage

\subsection{Model complexity}
\label{sec:model_com}

The computational cost stands as another crucial factor in model evaluation. Table \ref{tab:model_complexity} shows various metrics including the number of parameters, model size, training time for one epoch (the batch size of each model varies based on optimization results), inference time for one event (using a batch size of 256), and electron background rejection $R_e$ ($\varepsilon_{\pi}$ = 99\%) for each ANN-based model, all computed using a single NVIDIA V100 NVLink GPU. 

While our ResNet-based classifier and its upgraded version, the DGRes designed for point cloud scenarios, have an order of magnitude more parameters than DGCNN, GravNet, and LeNet, they demonstrate superior performance in electron background rejection in image and graph cases respectively. This improved performance is achieved without resulting in a proportional increase in the computational cost. Furthermore, the ResNet model based on image data exhibits a quite short inference time (1.45 ms) compared with graph-based approaches, which can be attributed to the benefits of the optimized parallel processing capabilities of the GPU operating on tensors. In contrast, the inference time of the graph-based networks in our study is influenced by the designated number of vertices in total and the nearest neighboring vertices. This is due to the necessity of sequentially covering all vertices to dynamically establish edge connections and features. Achieving parallel processing of vertices could enhance the efficiency of this dynamic construction process. The relatively higher time cost associated with DGRes and GravNet can be attributed to their requirement for a greater number of vertices (256) to construct a sub-graph during each iteration, based on our optimization findings. 

\begin{table}[h]
    \renewcommand{\arraystretch}{1}
    \caption{Number of parameters, model size, electron background rejection ($\varepsilon$ = 99\%, uncertainty is not listed), training time for one epoch, and inference time for one event of different models. Training and inference time are recorded using a single NVIDIA V100 NVLink GPU. For training, DGCNN employs a batch size of 128, GravNet utilizes a batch size of 32, AlexNet takes a batch size of 32, LeNet employs a batch size of 256, DGRes adopts a batch size of 32, while ResNet employs a batch size of 64, with optimal performance in background rejection. Inference time for all models is computed using a batch size of 256.}
    \centering
    \label{tab:model_complexity}
       
    \begin{tabular}{lccccc}
    \toprule
     & Parameters & Model size & $R_{e}^{\varepsilon=99\%}$ & Training [s/epoch] & Inference [ms/event] \\
    \hline

    DGCNN &  $1.8\times10^6$ & 6.87 MB & 2772 & \textbf{99.96} & 2.69  \\
    
    GravNet & $3.4\times10^6$ & 13.02 MB & 2516 & 624.11 & 12.60  \\
    
    AlexNet & $5.7\times10^7$ & 217.48 MB & 1864 & 51.55 & 1.56  \\
    
    LeNet & $6.7\times10^6$ & 25.41 MB & 1514 & \textbf{13.67} & \textbf{0.71}  \\
    
    DGRes & $1.4\times10^7$ &  53.92 MB & \textbf{3758} &  338.31 & 4.89  \\
    
    ResNet & $1.4\times10^7$ &  53.66 MB & 2165 &  105.41 &  1.45  \\
    
    \bottomrule
    \end{tabular}
\end{table}

\begin{figure}[h]
\centering

\includegraphics[width=0.8\textwidth]{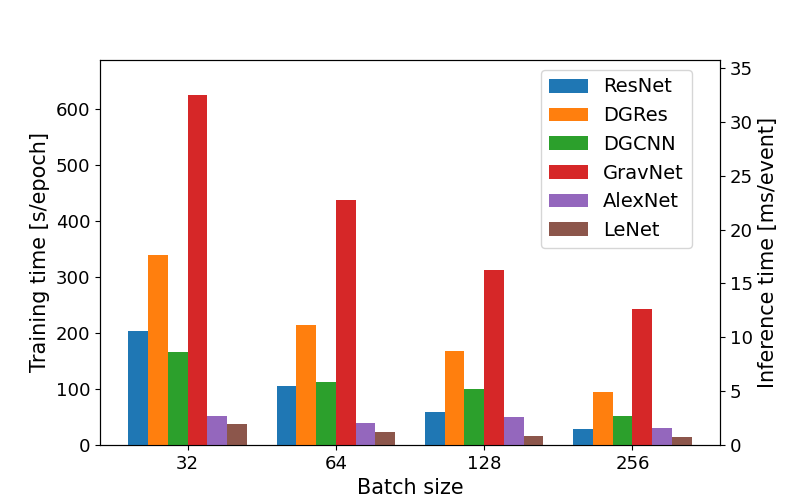}

\caption{The training and inference time with a variation of batch size.}
\label{fig:time}
\end{figure}

The training and inference performance of ResNet, DGRes, DGCNN, and GravNet with batch sizes ranging from 32 to 256 is further investigated as shown in figure \ref{fig:time}. Here, the inference time is calculated by event. Notably, we observe that ResNet (image-based) experiences the most significant benefit from the optimized parallel processing capabilities of the GPU. This is evidenced by the time cost for training and inference gradually becoming the lowest compared with other models when the batch size keeps increasing to 256, consistent with the feature of parallel processing. Consequently, this presents a disadvantage, particularly during large-batch inference in applications.

\subsection{Discussion}

Firstly, ANN provides a powerful framework for learning from data, with CNN excelling in processing grid-like data such as images, and GNN being specialized for handling graph-structured data as we observe that all ANN classifiers (ResNet, DGRes, DGCNN, GravNet, LeNet, ALexNet) outperform our BDT classifier in pion identification, which should be attributed to the advantage of ANN in harnessing the rich information present in high-dimensional inputs. Conversely, the intricate nature of electromagnetic and hadronic showers poses challenges for TMVA BDT, as it struggles to capture the complete shower details with a limited set of feature variables. Applying BDT is more akin to employing human-like thinking since it depends on pre-reconstructed human-readable variables. It is possible that our BDT will be further improved if more shower topology variables are reconstructed, but conducting this feature engineering would sometimes be tricky. 

Secondly, when applying ANN classifiers, both image-based and graph-based representations offer distinct advantages. In the case of the image-based representation, image data can be high-dimensional, resulting in increased computational complexity and memory requirements, particularly for large datasets. Moreover, image-based classifiers may encounter challenges in generalizing well to variations in the geometry of detectors. However, they can effectively leverage the parallel processing capabilities of GPUs, enhancing computational efficiency. Conversely, graph-based data offers enhanced flexibility in accommodating varying numbers of features and adapting to changes in the geometry of detectors. However, analyzing graph-based data can be computationally demanding, particularly for large graphs containing numerous nodes and edges. This complexity arises from the necessity to propagate information throughout the entire graph structure.

\section{Conclusion}
\label{sec:conclusion}

Novel PID methods based on Residual Artificial Neural Networks (ResNet) architecture are proposed to classify electron and hadron events based on the shower topology in high-granularity calorimeters, whose effectiveness is validated by using Monte Carlo simulation samples.

When classifying complex varieties of electromagnetic showers and hadronic showers, the introduction of the Residual Connection shortcut improves the robustness of neural networks. It also leads to superior performance, demonstrated in signal efficiency and background rejection, when image-based ResNet is compared with the BDT classifier, which relies on a limited number of input variables and still has room for improvement when additional relevant variables are deliberately developed. We also offer a model called DGRes, an upgrade to ResNet, by leveraging multiple EdgeConv blocks to address the application limitation when the tile size of the detector is not fixed. In this scenario, data can initially be represented in point clouds (graph-based), allowing for flexibility regardless of the detector's geometry. When compared with other state-of-the-art classifiers such as DGCNN and GravNet, our approach consistently demonstrates superior performance in electron background rejection.

As a result, our PID methods are practical, robust, feasible, efficient, and reliable in various imaging detectors with a huge amount of spatial information.

\appendix
\section{BDT input variables}
\label{ap:bdt}

\begin{figure}[htbp]
\centering
\subfigure[]
{   \label{sub:shw_den}
      
    \includegraphics[width=0.3\textwidth]{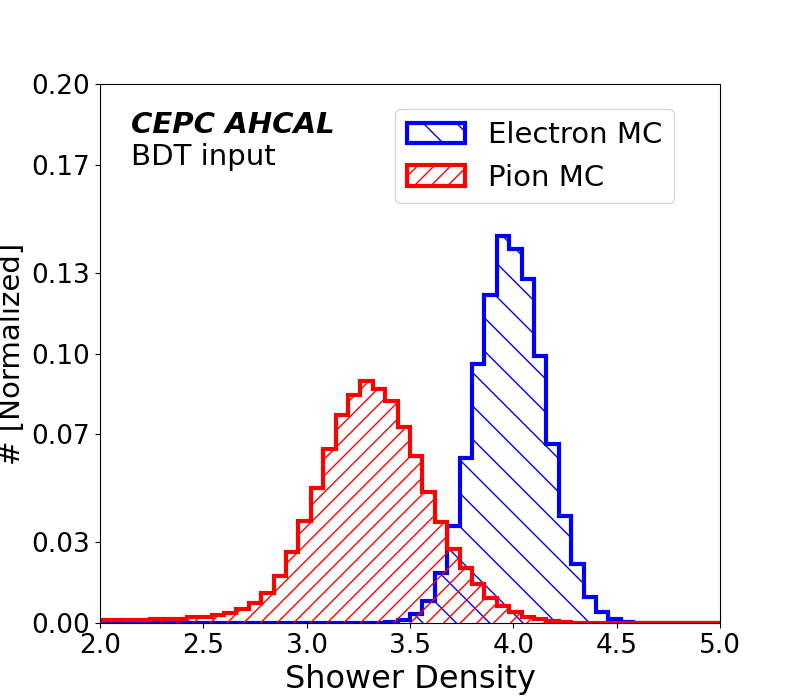}
           
}
\subfigure[]
{   \label{sub:shw_sta}
         
    \includegraphics[width=0.3\textwidth]{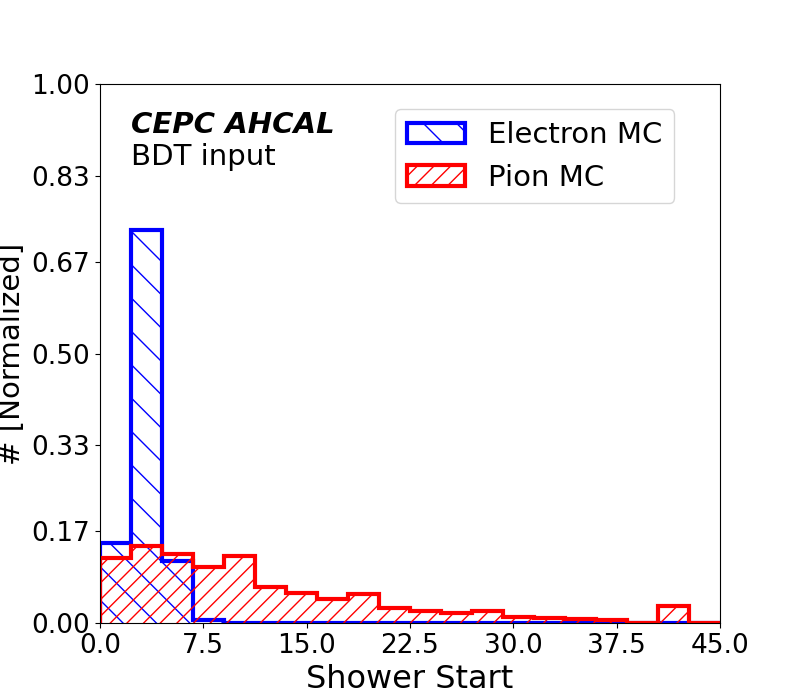}
    
}
\subfigure[]
{   \label{sub:shw_end}
 	
    \includegraphics[width=0.3\textwidth]{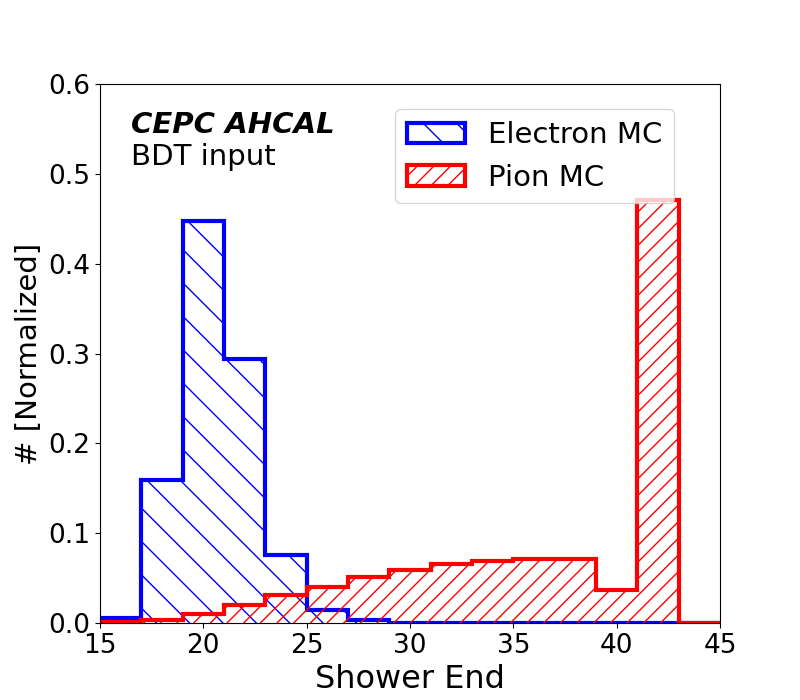}
   
}
\subfigure[]
{   \label{sub:shw_lg}
    
    \includegraphics[width=0.3\textwidth]{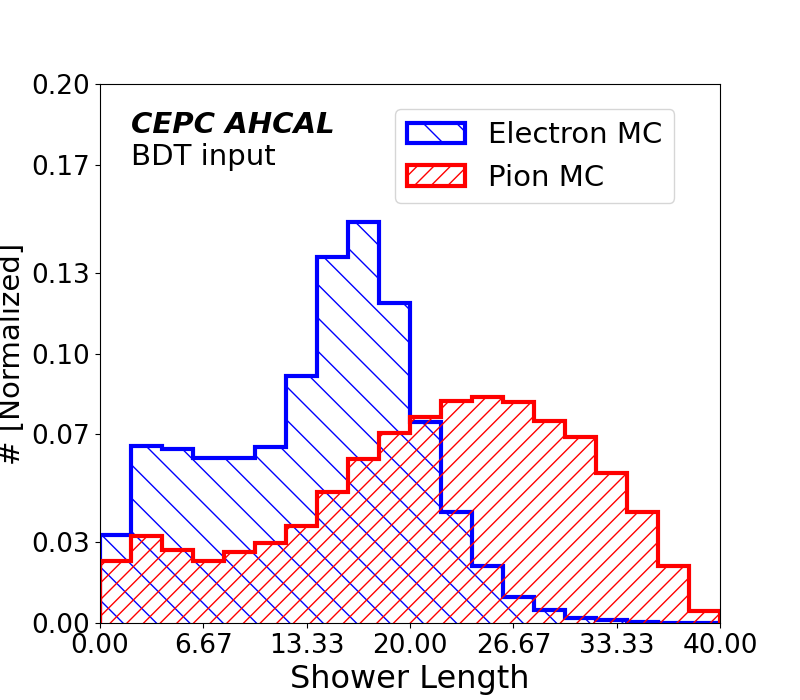}
   
}
\subfigure[]
{   \label{sub:h_n}
    
    \includegraphics[width=0.3\textwidth]{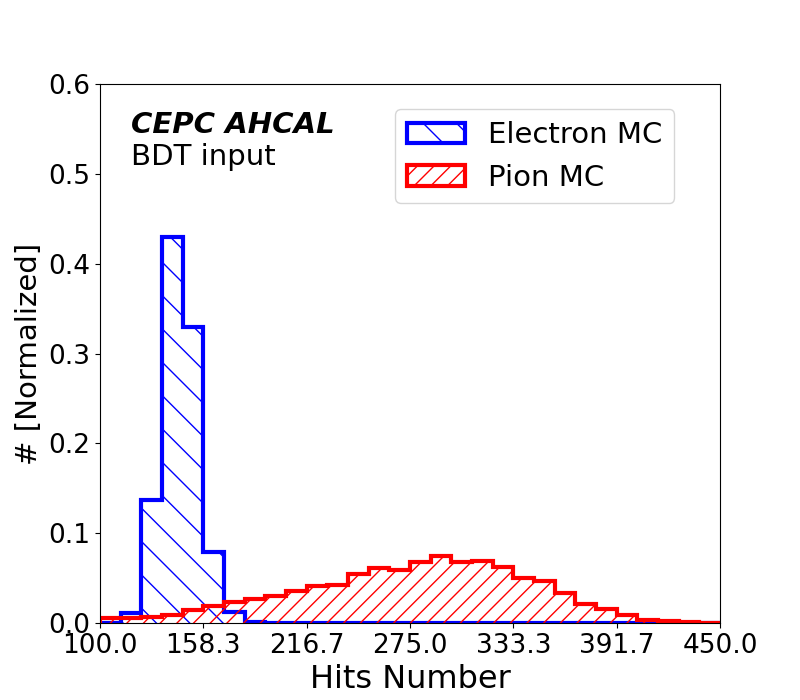}
 
}
\subfigure[]
{   \label{sub:shw_ra}
    
    \includegraphics[width=0.3\textwidth]{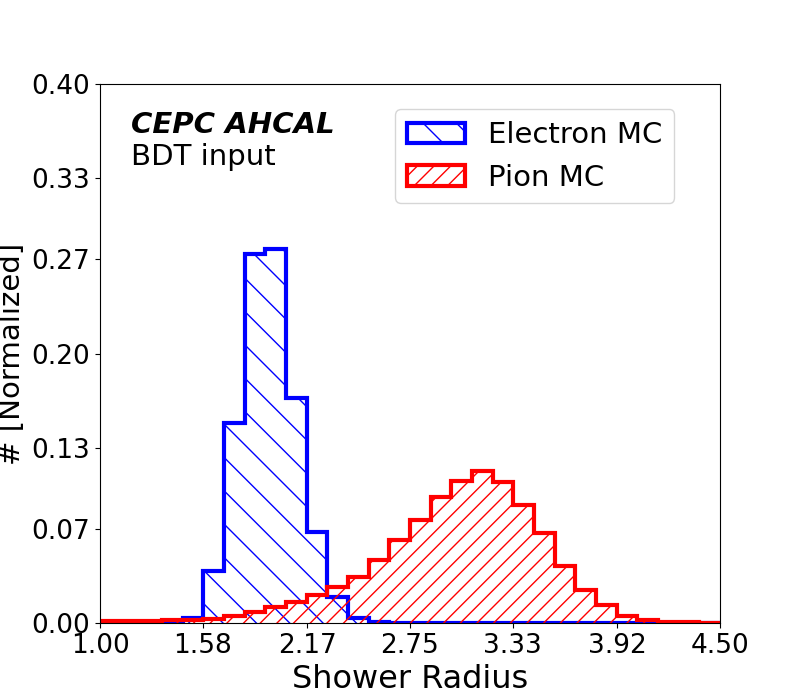}
   
}
\subfigure[]
{   \label{sub:fd_1}
 	
    \includegraphics[width=0.3\textwidth]{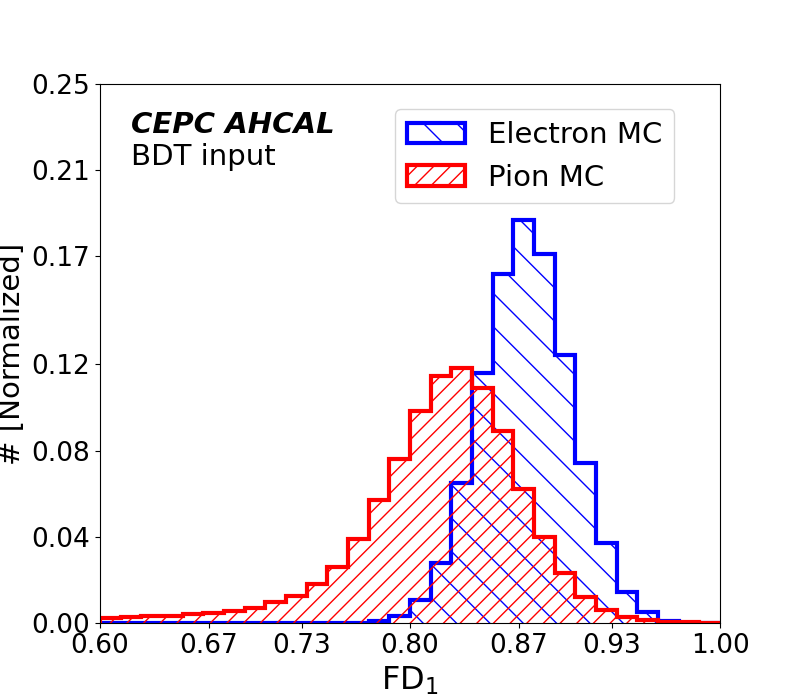}
   
}
\subfigure[]
{   \label{sub:fd_6}
 	
    \includegraphics[width=0.3\textwidth]{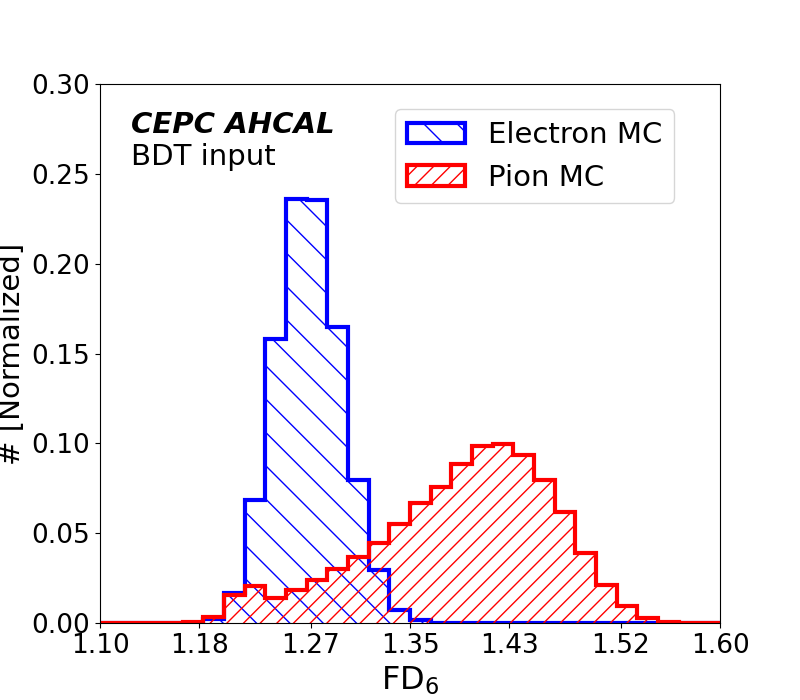}
   
}
\subfigure[]
{   \label{sub:fir_layer}
           
    \includegraphics[width=0.3\textwidth]{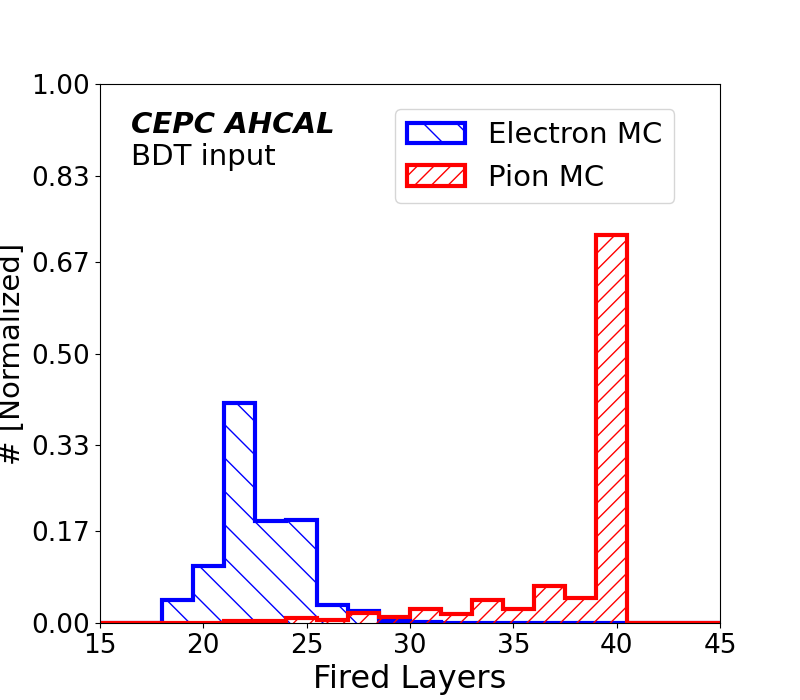}
   
}
\subfigure[]
{   \label{sub:shw_layer}
 	
    \includegraphics[width=0.3\textwidth]{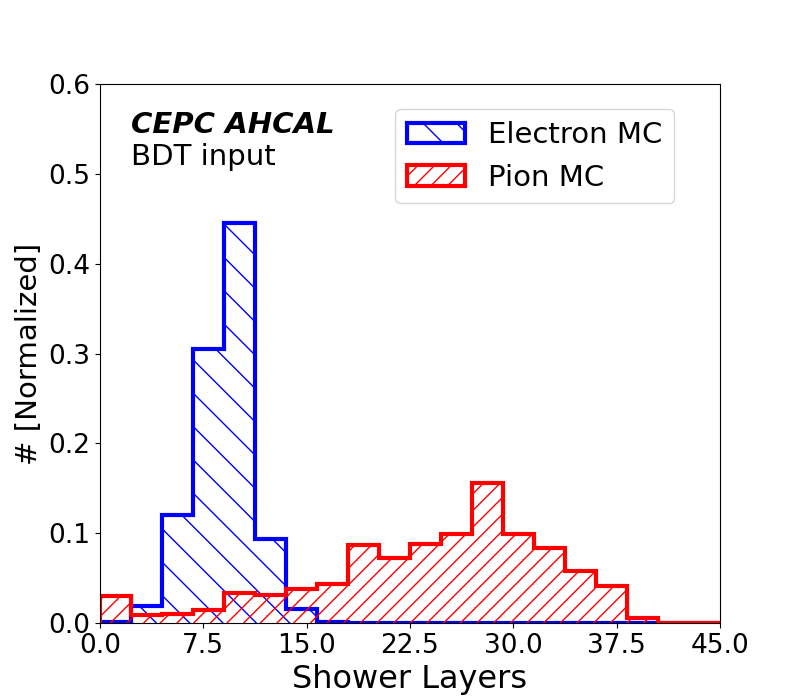}
   
}
\subfigure[]
{   \label{sub:shw_lr}
           
    \includegraphics[width=0.3\textwidth]{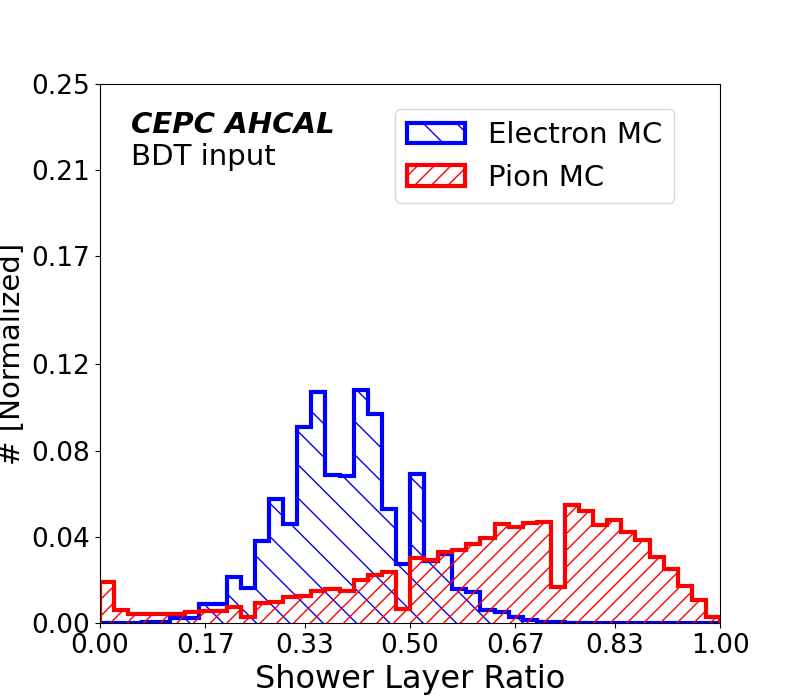}
   
}
\subfigure[]
{   \label{sub:z_w}
 	
    \includegraphics[width=0.3\textwidth]{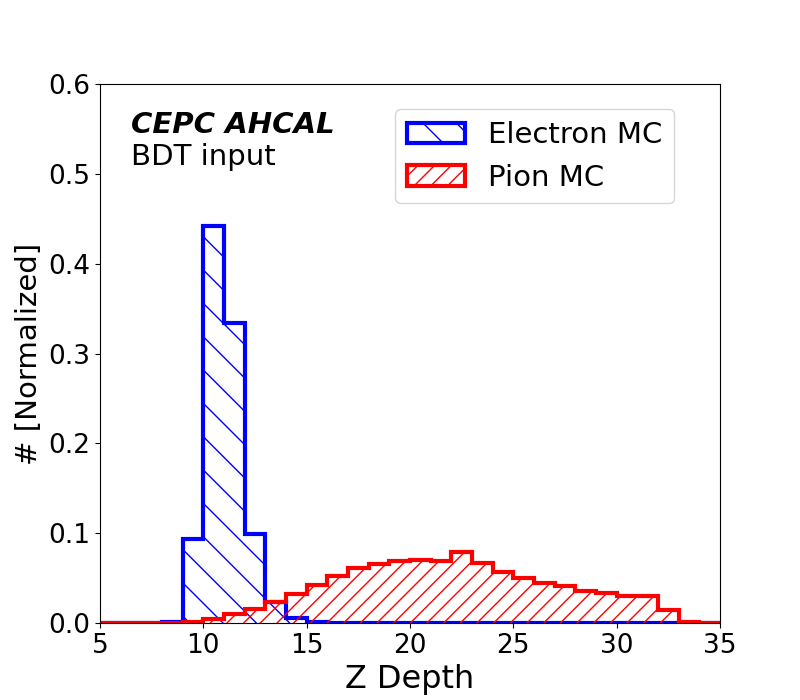}
   
}
\caption{The distribution of the Shower Density (a), the Shower Start (b),  the Shower End (c), the Shower Length (d), the Hits Number (e), the Shower Radius (f), the $\mathrm{FD}_1$ (g), the $\mathrm{FD}_6$ (h), the Fired Layers (i), the Shower Layers (j), the Shower Layer Ratio (k), and the Z Depth (l).}
\label{fig:bdt_var}   
\end{figure}

In the AHCAL prototype, a conventional right-handed coordinate system is utilized, with its 40 layers arranged perpendicular to the incoming beams. The origin of this coordinate system is established at the center of the first layer. The transverse plane, referred to as the $x$--$y$ plane, is parallel to the AHCAL layers. The $z$-axis is aligned with the direction of the incoming beam. 12 commonly utilized shower topology variables are reconstructed from the spatial distribution of energy depositions in the AHCAL prototype:

\begin{itemize}
    \item \textbf{Shower Density}: The average number of neighboring hits around one hit, including the hit itself, in a $3\times3$ cell area in a given event is calculated. The distributions of this variable, referred to as the shower density, are shown in figure \ref{sub:shw_den}. As expected, electromagnetic showers exhibit a more compact distribution compared to hadronic showers.
    
    \item \textbf{Shower Start}: The first layer of the first three consecutive layers with at least 5 hits. For events without showers, the shower start layer is set to 42. Electrons start showering virtually in the first 7 layers as illustrated in figure \ref{sub:shw_sta}.

    \item \textbf{Shower End}: After the shower starts, the first layer of two consecutive layers with no more than 2 hits. If no shower is formed, it is set to 42. It is illustrated in figure \ref{sub:shw_end} that electromagnetic showers almost end in the first half of the AHCAL.

    \item \textbf{Shower Length}: The distance between the start of the shower and the layer where the maximum Root Mean Square (RMS) of hit transverse coordinates concerning the $z$-axis occurs. In figure \ref{sub:shw_lg}, it can be observed that some pion events exhibit a longer shower length.
    
    \item \textbf{The total number of hits (Hits Number)}: In a given event, it represents the total number of hits. As shown in figure \ref{sub:h_n}, It can effectively separate electromagnetic showers and hadronic showers.

    \item \textbf{Shower Radius}: The RMS of the distance with respect to the $z$-axis of AHCAL. Figure \ref{sub:shw_ra} illustrates that electrons and pions exhibit distinct values in this variable.

    \item \textbf{Fractal dimension (FD)}: Fractal dimension only depends on the calorimeter observables \cite{ruan2014fractal}. By grouping blocks of $\alpha\times\alpha$ cells, where $\alpha$ defines the scale at which the shower is analyzed, $\mathrm{N}_{\alpha}$, as the number of hits at scale $\alpha$, could be calculated. Then, varying a series of $\alpha$ larger than $\beta$, $\mathrm{FD}_{\beta}$ could be derived based on equation \ref{eq:fd}. As shown in figures \ref{sub:fd_1} and \ref{sub:fd_6}, pions and electrons exhibit distinct values.

    \begin{equation}
    \mathrm{FD}_{\beta}=\left \langle {\frac{\mathrm{log(R_{\alpha,\beta})}}{\mathrm{log(\alpha)}}} \right \rangle+1,\ \mathrm{where }\ \mathrm{R}_{\alpha,\beta}=\mathrm{N}_\beta/\mathrm{N}_\alpha
    \label{eq:fd}
    \end{equation}

    \item \textbf{The number of layers with hits (Fired Layers)}: It is defined as the number of layers with at least one hit. In figure \ref{sub:fir_layer}, we can find that this variable helps to discriminate electron events from pion events.
    
    \item \textbf{Shower layers}: The number of layers in which the RMS of positions in the $x$--$y$ plane exceeds $4$ cm in both the $x$ and $y$ directions. Figure \ref{sub:shw_layer} illustrates that electrons and pions exhibit distinct values in this variable.

    \item \textbf{Ratio of shower layers over fired layers (Shower Layer Ratio)}: This variable has differences between electrons and pions, as depicted in figure \ref{sub:shw_lr}.

    \item \textbf{Z Depth}: The RMS of the $z$-axis coordinates. There is a clear difference between electron events and pion events in this variable, as shown in figure \ref{sub:z_w}.

\end{itemize}

\newpage

\section{Training error and validation error}
\label{ap:error}

\begin{figure}[htbp]
\centering

\subfigure[]
{   \label{sub:err_64_00001}    
    \includegraphics[width=0.3\textwidth]{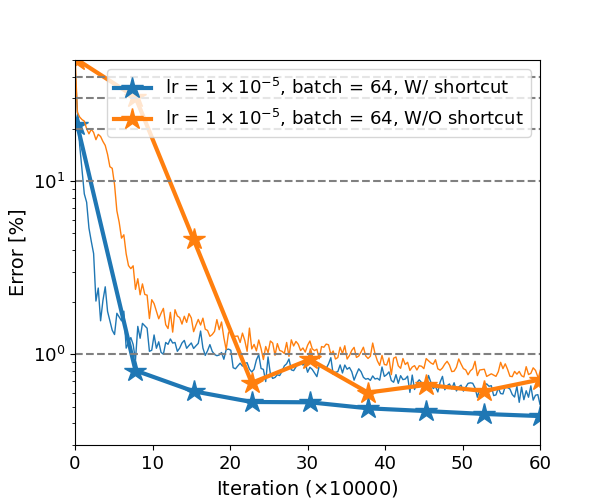}   
}
\subfigure[]{ 
    \label{sub:err_128_00001}
    \includegraphics[width=0.3\textwidth]{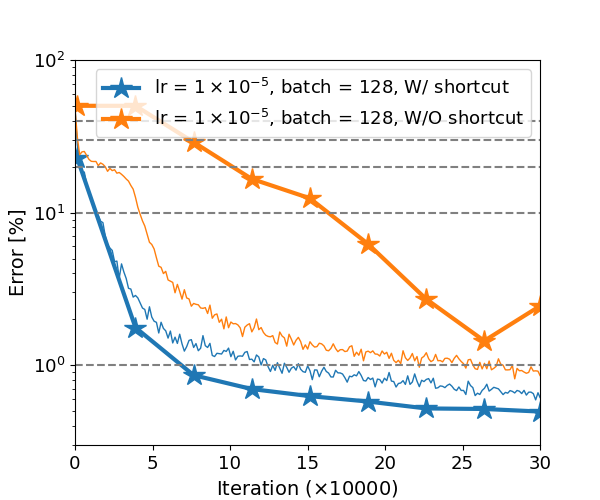}
}
\subfigure[]{ 
    \label{sub:err_256_00001}
    \includegraphics[width=0.3\textwidth]{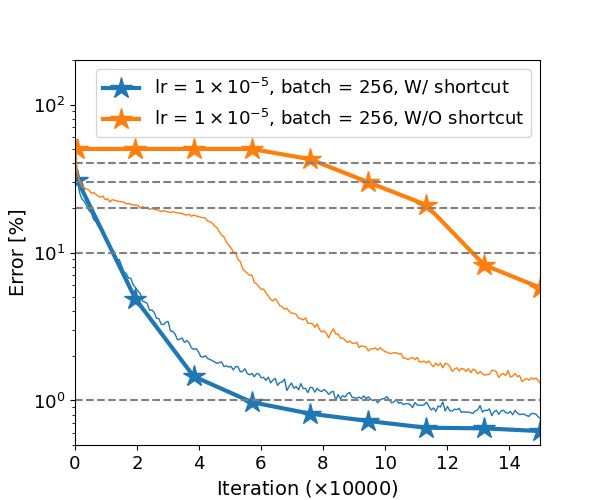}
}
\subfigure[]
{   \label{sub:err_64_0001}    
    \includegraphics[width=0.3\textwidth]{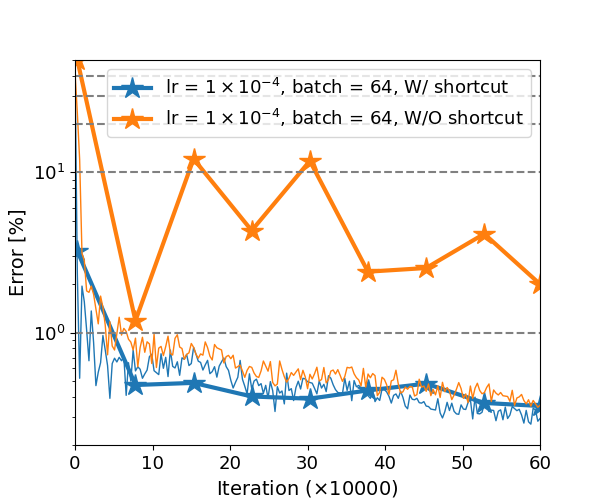}   
}
\subfigure[]{ 
    \label{sub:err_128_0001}
    \includegraphics[width=0.3\textwidth]{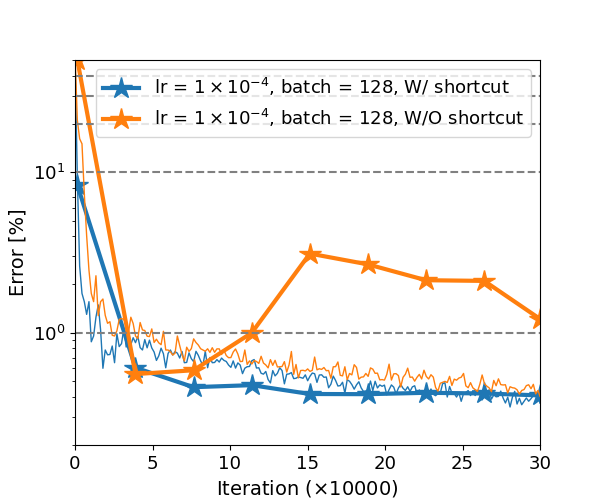}
}
\subfigure[]{ 
    \label{sub:err_256_0001}
    \includegraphics[width=0.3\textwidth]{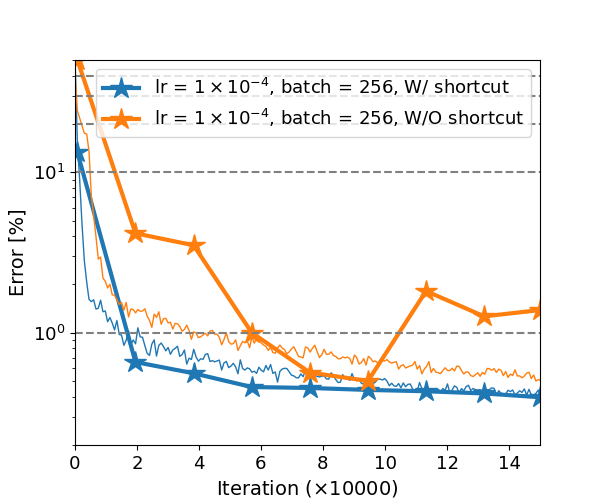}
}
\subfigure[]
{   \label{sub:err_64_001}    
    \includegraphics[width=0.3\textwidth]{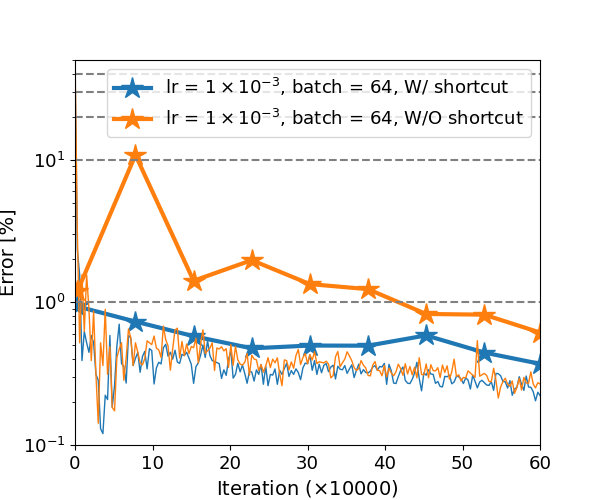}   
}
\subfigure[]{ 
    \label{sub:err_128_001}
    \includegraphics[width=0.3\textwidth]{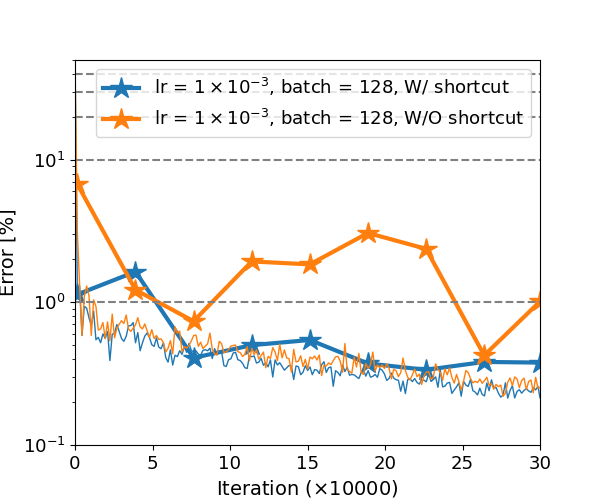}
}
\subfigure[]{ 
    \label{sub:err_256_001}
    \includegraphics[width=0.3\textwidth]{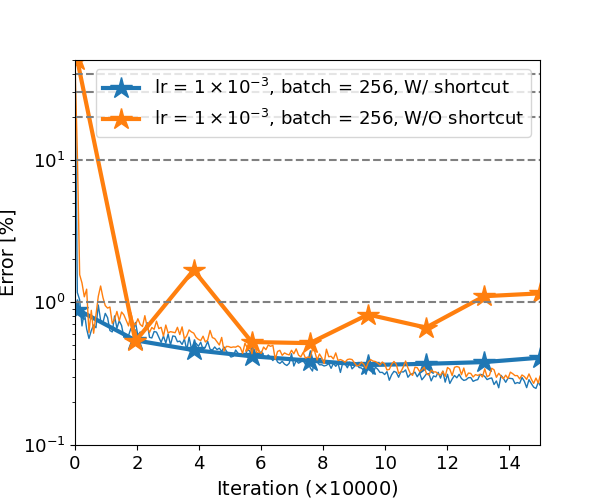}
}

\caption{Training the ResNet classifier involves adjusting the learning rate ($lr$), set at $1\times10^{-5}$ in the first row, $1\times10^{-4}$ in the second row, and $1\times10^{-3}$ in the third row. Additionally, the batch size varies across three configurations: 64 (left), 128 (middle), and 256 (right). In the plot, thin curves represent the training error, while bold curves depict the validation error. The training error is updated after each epoch, and the validation error is updated every 25 epochs using the independent validation set. Notably, the only distinction in the same plot lies in the presence or absence of shortcuts.}
\label{fig:error_shortcut}
\end{figure}

Figure \ref{fig:error_shortcut} illustrates the impact of the shortcut connection on the training error and the validation error during the iteration across various combinations of the batch size (64,~ 128,~ 256~) and the learning rate ($1\times10^{-5}$, $1\times10^{-4}$, $1\times10^{-3}$) while maintaining 200 epochs.

\acknowledgments

We thank the CEPC calorimeter group and the CALICE Collaboration for their support and assistance. The study was supported by National Natural Science Foundation of China (No.11961141006), National Key Programmes for S\&T Research and Development (Grant No.: 2018YFA0404300 and 2018YFA0404303).



\bibliographystyle{JHEP}
\bibliography{biblio.bib}

\end{document}